\documentclass{article}
\usepackage[numbers]{natbib}
\usepackage[preprint]{neurips_2025}

\usepackage[utf8]{inputenc} 
\usepackage[T1]{fontenc}    
\usepackage{hyperref}       
\usepackage{url}            
\usepackage{booktabs}       
\usepackage{amsfonts}       
\usepackage{nicefrac}       
\usepackage{microtype}      
\usepackage{xcolor}         
\usepackage{graphicx}
\usepackage{amsmath, amsthm, amssymb}

\usepackage{wrapfig}
\usepackage{etoolbox} 
\AtBeginEnvironment{proof}{%
  \setlength{\topsep}{0pt}%
  \setlength{\partopsep}{0pt}%
}

\title{Predicting Multi-Agent Specialization  \\ via Task Parallelizability}

\author{
  \bfseries Elizabeth Mieczkowski\textsuperscript{1}\footnotemark[1] \quad
  \bfseries Ruaridh Mon-Williams\textsuperscript{2}\footnotemark[1] \quad
  \bfseries Neil Bramley\textsuperscript{2}\footnotemark[2] \\
  \bfseries Christopher G.\ Lucas\textsuperscript{2}\footnotemark[2] \quad
  \bfseries Natalia V\'elez\textsuperscript{1}\footnotemark[2] \quad
  \bfseries Thomas L.\ Griffiths\textsuperscript{1}\footnotemark[2] \\
  \textsuperscript{1}Princeton University \quad
  \textsuperscript{2}University of Edinburgh \\
  \texttt{emiecz@princeton.edu, ruaridh.mw@ed.ac.uk}
}

\begin{document}

\maketitle

\footnotetext[1]{Equal contribution.}
\footnotetext[2]{Equal senior authorship.}

\begin{abstract}

    When should we encourage specialization in multi-agent systems versus train generalists that perform the entire task independently? 
    We propose that specialization largely depends on task parallelizability: the potential for multiple agents to execute task components concurrently. 
    Drawing inspiration from Amdahl's Law in distributed systems, we present a closed-form bound that predicts when specialization improves performance, depending only on task concurrency and team size. We validate our model on two standard MARL benchmarks that represent opposite regimes--- StarCraft Multi-Agent Challenge (SMAC, unlimited concurrency) and Multi-Particle Environment (MPE, unit-capacity bottlenecks)---and observe close alignment between the bound at each extreme and an empirical measure of specialization. Three follow-up experiments in Overcooked-AI demonstrate that the model works in environments with more complex spatial and resource bottlenecks that allow for a range of strategies. Beyond prediction, the bound also serves as a diagnostic tool, highlighting biases in MARL training algorithms that cause sub-optimal convergence to specialist strategies with larger state spaces. 
\end{abstract}

\section{Introduction}
\label{introduction}

Collaboration enables multi-agent systems to achieve goals that exceed the capabilities of a single agent. Advances in artificial intelligence (AI) have enhanced the capabilities of multi-agent systems to complete complex, coordinated tasks, including automating warehouses \cite{canese2021multi}, sustainable resource management \cite{perolat2017multi}, and simulated social interactions \cite{park2023generative}. The prevailing assumption in multi-agent reinforcement learning (MARL), cognitive science, and robotics is that \textbf{specialization} is universally desirable \cite{padgham2002prometheus, zhu2008role, charakorn2020investigating, wang2020roma, wang2020rode, li2021celebrating, bettini2023heterogeneous, juang2024breaking, bettini2024controlling, swanson2024virtual}. For example, fine dining restaurants rely on \textit{specialists}---the saucier prepares sauces, the r\^otisseur roasts, and so on---who each perform distinct yet complementary subtasks. Intuitively, specialization narrows action spaces, reduces ambiguity about what agents should do next \cite{goldstone2024emergence}, and eases individual cognitive loads \cite{griffiths2020understanding}. 
Recent algorithms such as ROMA and RODE rely on this assumption to treat task allocation as an optimization problem, focusing on learning roles and constraining agents to specific subtasks \cite{wang2020rode, wang2020roma}. However, in both human and MARL teams, specialization introduces trade-offs. Fine dining chefs produce complex meals via rigid role divisions, but the system is brittle: training costs are high, and service halts if one specialist fails or takes longer than the others \cite{ben2023cultural, velez2024rise}. In contrast, generalist fry cooks, who all grill, assemble, and plate, maximize throughput by performing subtasks in parallel. While MARL agents do not fatigue or quit like humans, they do face coordination challenges, policy divergence, and retraining costs when roles are overly fixed \cite{shoham2008multiagent, wu2021too}. This raises a key question: what task and environmental constraints make specialization truly optimal, and when are generalists better?


\begin{figure*}[t]
     \centering
     \includegraphics[width = \linewidth]{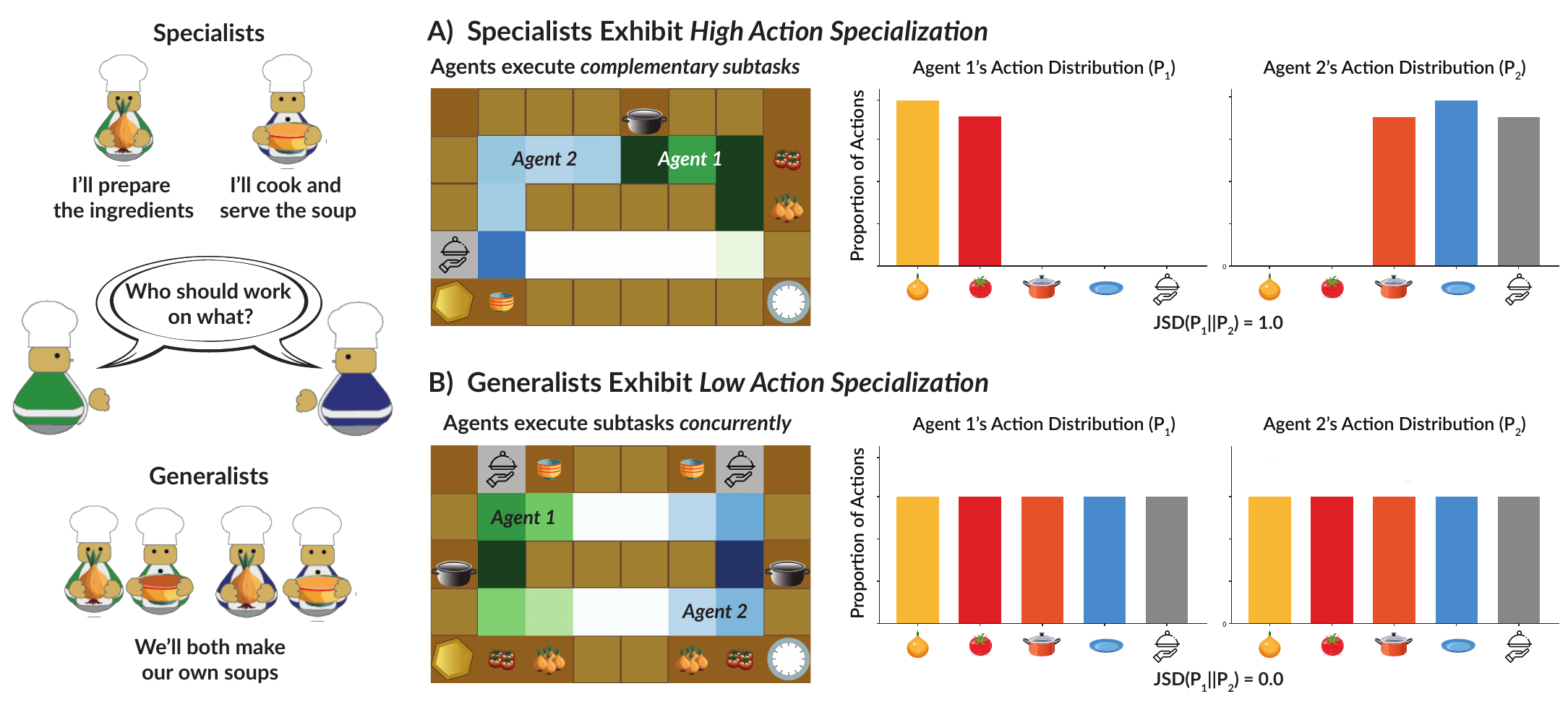}
    \caption{Specialists vs. generalists. (A) Specialist teams focus on non-overlapping subtasks, resulting in high Jensen-Shannon divergence (SI) between agents' action distributions. \href{https://drive.google.com/file/d/1s6sf4GXRDifk08IWHj4RyA5Awe8oDfI_/view?usp=sharing}{Example video.} (B) In generalist teams, both agents perform the task independently, resulting in low SI. \href{https://drive.google.com/file/d/1ORq1bQUcahpEq_1CPI59LrrIr0SF8wdf/view?usp=drive_link}{Video.} }
    \label{fig:SI}
\end{figure*}

We propose that the degree to which a task can be executed in parallel (or its \textbf{parallelizability}) directly governs the effectiveness of generalist teams. When spatial or resource bottlenecks force agents to wait, the benefits of independently performing a task in parallel vanish and specialist policies become more efficient. Drawing from distributed computing \cite{almasi1994highly, mccool2012structured}, we adapt Amdahl's Law to treat agents as processors and subtasks as units of work that may or may not be parallelizable \cite{amdahl1967validity, hill2008amdahl, cassidy2011beyond}. This yields a simple analytic bound on the speed-up achievable by generalist teams as a function of task structure. When the parallelizability of the overall task falls below team size, specialization is advantageous. 

We show that a task’s parallelizability provides a simple prediction for when specialist policies will emerge in multi-agent systems given a team size and environment. In Experiment 1, we apply the model to two edge-case MARL benchmarks: SMAC, where generalists dominate due to unlimited concurrency, and MPE, where unit-capacity bottlenecks force agents to specialize. In Experiments 2 and 3, we extend to more complex spatial and resource dynamics in Overcooked-AI, demonstrating a more fine-grained relationship between task parallelizability and specialization. Experiment 4 shows that larger state spaces promote deviations from model predictions, highlighting biases in existing MARL training regimes. Together, these findings suggest that specialization is not obligatory for successful teams, and is affected by bottlenecks and coordination structure. Our framework enables the design of environments that encourage or inhibit policy diversity and illustrates how tools from distributed systems can illuminate multi-agent settings.

\section{Task Parallelizability Framework}
\label{amdahl}
In many multi-agent settings, the marginal benefit of adding agents is limited not only by training but also \emph{structural constraints} on the task and environment. Quantifying task parallelizability answers two key questions: (1) When should we encourage behavioral diversity and specialization during training, and (2) When parallelism is limited, which \emph{dimension}— team size $N$,
subtask allocation $f_i$, spatial capacity $C^s$, or resource capacity $C^r$— is responsible for preventing speed-up? The key result (§2.4) shows that when overall parallelizability $S(N,B)$ drops below team size, specialization becomes strictly more efficient. Sections 3–5 demonstrate how to measure concurrency constraints for various MARL applications and validate the prediction empirically. 

\subsection{$N$-Player Markov Game}
Following \citet{albrecht2024multi}, 
we formalize a cooperative decentralized Markov decision process (Dec-MDP), defined as $\mathcal{M} = \langle \mathcal{S}, \mathcal{N}, A, \mathcal{T}, \mathcal{R}, \gamma \rangle $. $\mathcal{S}$ is a finite set of states, and $\mathcal{N}$ is a set of $N$ agents. $A$ is the joint action space, where each agent $i \in \mathcal{N}$ selects an action $a_i \in A$ to form a joint action $a = (a_1,a_2, ..., a_N)$. $\mathcal{T} : \mathcal{S} \times A \to \Delta(\mathcal{S})$ is the transition function which defines the probability distribution over next states given the current state and joint action. Finally, $\mathcal{R} : \mathcal{S} \times A \to \mathbb{R}$ is the shared reward function. Each agent's objective is to maximize its expected discounted return $\mathbb{E}\big[\sum_{t=0}^{\infty}\gamma^t r(s_t,a_t)\big]$, where $\gamma \in [0,1)$ is a discount factor and $r(s_t,a_t)$ is the shared reward received by all agents at time $t$. Agents share the same reward but optimize their returns independently. 

\subsection{Jensen-Shannon Divergence and Specialization}
\label{SI}
We define \textbf{specialization} as the differentiation of agents' policies, capabilities, or roles in a multi-agent task. Let $\pi^i : \mathcal{S} \to \Delta(\mathcal{A}^i)$ be the policy of agent $i$ that maps each state $s \in \mathcal{S}$ to a distribution over actions $a \in \mathcal{A}^i$. Each agent follows a state-visitation distribution that describes how frequently they occupy different states and actions under $\pi^i$, which can be used to define their effective action distributions $P_i(a)$ \cite{schulman2015trust}. For a complete formulation, refer to Appendix~\ref{stateactionoccupancy}. 
We approximate specialization using the \textbf{Jensen-Shannon divergence (JSD)} over agents' action distributions $P_i$ \cite{endres2003new, fuglede2004jensen}. We use JS rather than Kullback-Leibler divergence because it is (a) well-defined when agents' action distributions have disjoint support and (b) symmetric ($\text{SI}(P_1||P_2) = \text{SI}(P_2||P_1)$), since the degree of specialization should be the same between teams regardless of the order of agents. For $N$ agents with action distributions $P_1, P_2, ..., P_N$, JSD is given by:
\begin{equation}
    \operatorname{JSD}(P_{1:N}) \;=\;
    \frac1N \sum_{i=1}^N
    D_{\mathrm{KL}}\!\bigl(P_i \,\|\, M\bigr),
    \qquad
    M=\frac1N \sum_{i=1}^N P_i,
\end{equation}
where $M$ is the mixture distribution and \(D_{\mathrm{KL}}(P_i||M)\) is the Kullback-Leibler (KL) divergence between each agent's action distribution $P_i$ and $M$. Because the maximal value grows as $\max\operatorname{JSD}(P_{1:N})=\log_2 N$ bits, we report the
\textbf{Specialization Index}:

\begin{equation}
    \mathrm{SI}(P_{1:N}) \;=\;
    \frac{\operatorname{JSD}(P_{1:N})}{\log_2 N}\;\in[0,1]
\end{equation}

throughout the paper. $\mathrm{SI}=0$ indicates identical distributions; $\mathrm{SI}=1$ indicates perfectly non‑overlapping distributions. In fully \textbf{generalist} teams, agents select from the entire set of actions $A$ to complete the task in parallel. 
In this case, all $N$ agents have performed the same proportions of actions, leading to $\mathrm{SI}(P_1, P_2, ..., P_N)=0.0$ as an indicator of no specialization. Alternatively, in fully \textbf{specialist} teams, 
the action distributions of each agent are completely disjoint (i.e., each agent assigns probability exclusively to a unique set of actions), meaning that $\mathrm{SI}(P_1, P_2, ..., P_N)=1.0$ indicates complete specialization. Intermediate values of SI indicate partial specialization, in which the agents overlap in some subtasks or occupancies but not others. Examples of action distributions and specialist versus generalist strategies can be found in Figure \ref{fig:SI}. 

\subsection{Multi-Agent Bottlenecks}

In distributed computing systems, Amdahl's Law provides an upper bound on the speedup achievable through parallelism based on the proportion of a task that must be performed sequentially by a single processor \cite{amdahl1967validity, hill2008amdahl, cassidy2011beyond}. Certain computations inherently depend on centralized coordination or the results of previous steps
\cite{almasi1994highly, mccool2012structured, zhuravlev2010addressing}. A classic example is memory bandwidth in GPUs, where increasing the number of threads does not lead to proportional speed-up due to limited shared access \cite{dublish2017evaluating}. By extending Amdahl's Law to multi-agent systems, we can formulate a principled prediction for when tasks can be performed concurrently by agents or must be performed sequentially by one.
We begin by identifying \textbf{bottlenecks} that prevent parallelism in multi-agent tasks. 
Various task constraints have been identified across numerous multi-agent domains, such as warehouse operations (insufficient space, storage locations, and pallets \cite{vzivivcnjak2022case}), autonomous vehicles (lane reductions, traffic congestion \cite{zhang2020will}), and smart factories (equipment, production routes) \cite{wang2016towards}. Inspired by these findings, we account for two key types of bottlenecks: (1) \textbf{spatial bottlenecks} limit the number of agents ($C^s$) that can navigate due to obstacles, narrow passages, or congestion, and (2) \textbf{resource bottlenecks} limit accesses ($C^r$) that can be made at a time to critical resources such as workstations or tools \cite{mieczkowskipeople}. 

\paragraph{Task decomposition.}
Below, we offer an operationalization of bottlenecks that can be applied to a wide variety of tasks. Following \citet{solway2014optimal} and \citet{amato2019modeling}, we treat a cooperative task as a directed acyclic graph (DAG) \(T=(V,E)\). Each node \(i\in V\) is a subtask that can be executed by one or more agents, and an edge \((i\!\to\!j)\in E\) indicates that \(j\) cannot start until \(i\) completes. Let \(m=|V|\) and denote by \(f_i\) the expected fraction of total execution time spent in node \(i\). We require only \(\sum_{i=1}^{m}f_i=1\); concrete estimators for \(f_i\) are introduced per domain in §3.

\paragraph{Concurrency capacity.}
For every subtask $i$, let the concurrency capacity $C_i = \min\bigl(C^s_i,\,C^r_i\bigr)$ denote the maximum number of agents that can execute subtask $i$ in parallel before contention forces at least one agent to wait.
$C^s_i$ captures purely \emph{spatial} limits (e.g.\ a one-cell doorway), while $C^r_i$ captures \emph{resource} limits (e.g.\ two welding stations). We define subtasks independently of spatial location; two agents chopping vegetables on opposite ends of the room are treated as executing the same subtask.
Although the exact estimator is environment-specific
(see §3 for examples), most simulators support enumerating access paths or available resources and computing path widths. Alternatively, it is possible to count simultaneous agent accesses to each subtask during a short probe rollout. The minimum of these counts yields $C_i$. While we focus on spatial layout and resource access as primary limits of concurrency, other environments might impose additional bottlenecks, such as the latency required to exchange messages in communication-based environments. These bottlenecks, though not explored in this paper, can also be integrated into this concurrency score. 

\subsection{Task Parallelizability Prediction}
\label{finalprediction}
We now adapt Amdahl's Law to multi-agent tasks. For the original form, see Appendix~\ref{originalamdahl}. \citet{mieczkowskipeople} extended this idea to predict idleness during human collaborations given the current group size, workload, and resource constraints. We hypothesize that a similar formulation can be used to predict parallelizability in MARL. 
Let $f_i$ represent the fraction of the total task time required for the $i^{th}$ of $m$ subtasks such that $\sum_{i=1}^m f_i = 1$. For each subtask $i$, the overall bottleneck capacity is defined as the minimum bottleneck that prevents agents executing it concurrently, or $C_i = \min(C^s_i,C^r_i)$. The \textbf{subtask speed-up factor} $s_i(N,C_i)$ quantifies the efficiency gain from adding $N$ agents to a subtask, as opposed to a single agent performing it, given that $C_i$ bottlenecks restrict parallel execution. Formally, $s_i(N,C_i) = \text{min}(N,C_i)$. 
Thus, if we assume that there is no cost to switching between subtasks or coordinating with larger teams, the speed-up achievable when $N$ agents perform the entire task in parallel, which we call \textbf{task parallelizability}, is expressed as:
\begin{equation}
\label{speedup_equ}
    S(N,C) = \frac{1}{\sum_{i=1}^m \frac{f_i}{s_i(N,C_i)}}
\end{equation}

\paragraph{Properties} 
There are several useful properties of this speed-up bound $S(N, C)$ provided in Appendix~\ref{concurrencyproperties}. These propositions formalize the intuition that removing bottlenecks promotes generalists, never specialists, since only generalists benefit from concurrent access to subtasks.

\section{Applications to MARL Environments}
\label{applications}

The bound in §2 is abstract; it relies only on the subtask breakdown $f_i$ and the concurrency capacity $C_i$. Here we show how to measure these values in three representative classes of multi-agent benchmarks: grid-worlds (Overcooked-AI),
continuous environments (MPE), and high-throughput RTS scenarios (SMAC). In each domain, we define subtasks, estimate $f_i$, compute spatial and resource capacities, and plug them into Equation~\ref{speedup_equ}.

\paragraph{StarCraft Multi-Agent Challenge (SMAC)}
In SMAC \cite{de2020independent}, agents cooperate to defeat enemy units. Each agent observes a local feature vector and selects from discrete actions such as \texttt{left} or \texttt{attack enemy 1}. Let $m$ (subtasks) be the number of enemy units that agents must defeat to gain reward. We assume all subtasks require roughly equal effort and set $f_i = \frac{1}{m} \text{ } \forall i \in m$. Since all agents can move and shoot freely, and combat maps are open with no obstacles, there are no spatial or resource bottlenecks. We define $C_i = \infty \text{ } \forall i \in m$. Substituting into Equation~\ref{speedup_equ}, we obtain $S(N,B) = N$, or maximum parallelizability. According to Proposition 2, we would predict fully generalist strategies with near-zero specialization (SI $=0$).

\paragraph{Multi-Particle Environment (MPE)}
In MPE's Spread task \cite{lowe2017multi}, agents must each cover a landmark. Let $m$ be the number of landmarks, and assume $f_i = 1/m$. Since only one agent can occupy each landmark at a time, \( C_i = 1 \) for all $i$, yielding $S(N,B) = 1$. This low parallelizability should lead to full specialization (SI $=1$). 

\paragraph{Overcooked-AI}
In Overcooked-AI \cite{carroll2019utility, wu2021too}, agents must cook recipes by navigating a grid and coordinating at shared workstations. We represent each recipe as a directed acyclic graph (DAG) of subtasks. For example, preparing onion soup involves the sequence: \textit{Onion → Pot → Cook → Bowl → Serve} (with \( m = 4 \)), as shown in Figure~\ref{fig:DAG}.
To extract subtask fractions and bottlenecks, we model each kitchen layout as a graph $G = (\mathcal{V}, \mathcal{E})$, where nodes $\mathcal{V}$ represent spatial locations and workstations, and edges $\mathcal{E}$ represent valid movements. Each subtask $i$ corresponds to a path between workstations. We estimate each subtask fraction using normalized path lengths $f_i = d(i) / \sum_{j=1}^m d(j)$, 
where $d(i)$ is the length of the shortest path needed to complete subtask $i$. We assume a fixed cost of 10 steps for \textit{Cook}. We then compute concurrency limits per subtask. The spatial bottleneck $C^s_i$ is based on edge betweenness centrality along subtask $i$'s path, reflecting the narrowest point in available throughput \cite{barrat2004architecture}. The resource bottleneck $C^r_i$ is defined as the number of agents that can simultaneously use the workstation(s) required for subtask $i$. The overall concurrency limit is $C_i = \min(C^s_i, C^r_i)$. 
These values are substituted into Equation~\ref{speedup_equ} to compute parallelizability $S(N, B)$. 
For full computation details, see Appendix~\ref{layoutgraphs}.

\section{Experiment 1: Divergent Specialization Patterns Between Environments}

\begin{figure}
    \centering
    \begin{minipage}{0.65\textwidth}
     \includegraphics[width = \linewidth]{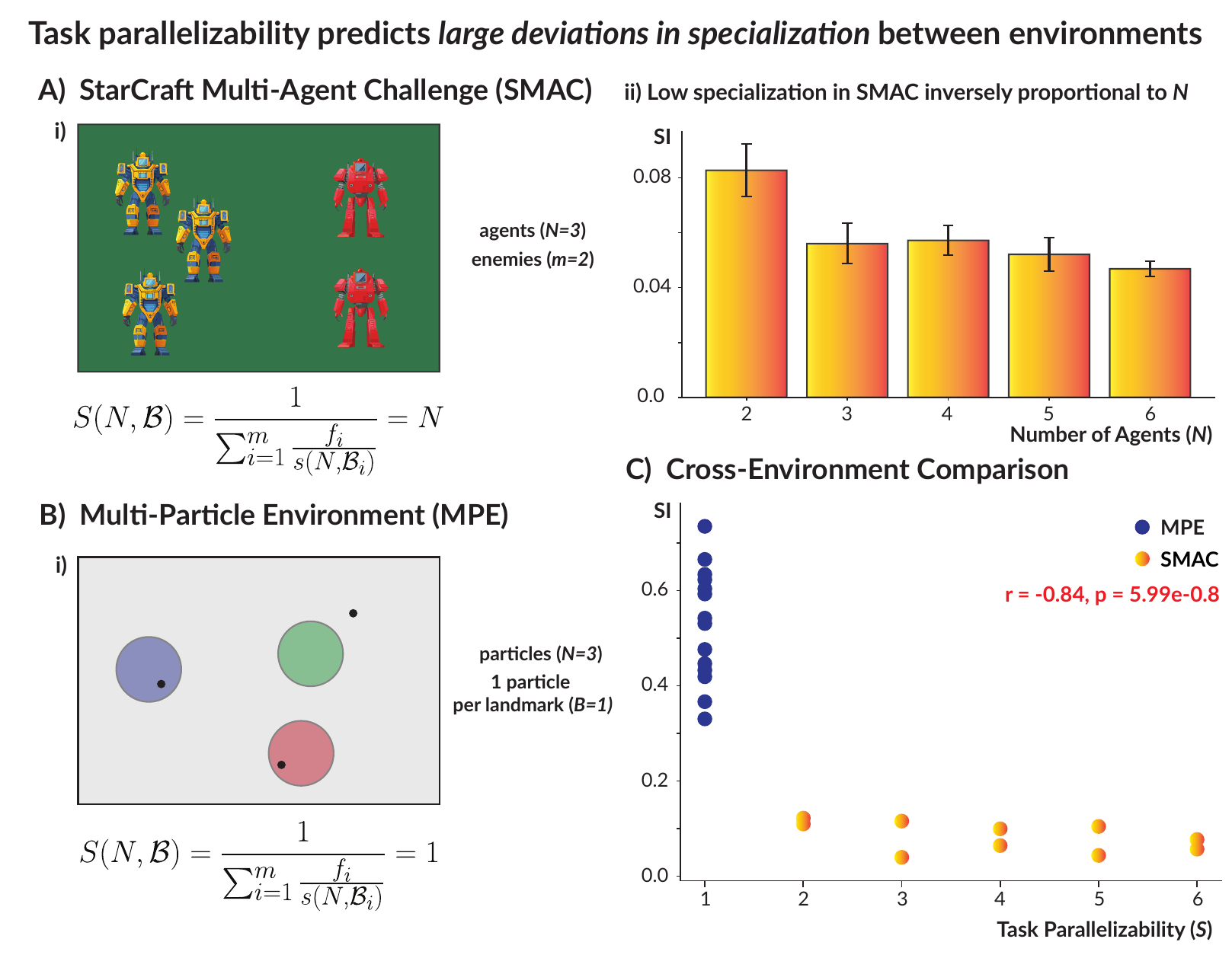}
     \end{minipage}
     \hfill
     \begin{minipage}{0.34\textwidth}
    \caption{Experiment 1 Results. (A) In SMAC, $N$ agents cooperate to defeat enemy units. (i) With no bottlenecks, parallelizability predicts $S(N, C) = N$. (ii) Specialization (SI) decreases with increasing parallelizability, as predicted ($r = -0.86$). (B) In MPE, agents cooperate to cover $m$ landmarks that can be occupied by 1 agent. (C) Across environments, task parallelizability $S$ is strongly inversely correlated with specialization ($r = -0.84$), and the logistic regression classifier achieves perfect accuracy ($\text{accuracy} = 1.0$).}
    \label{fig:smacmpe}
    \end{minipage}
\end{figure}

In Experiment 1, we test whether task parallelizability can predict deviations between specialist and generalist policies in vastly different environments, providing empirical validation for our framework. In SMAC, where agents face minimal spatial and resource constraints, generalist policies prevail and specialization further decreases with team size, consistent with high task parallelizability. In contrast, MPE imposes strict bottlenecks since each landmark can only be occupied by a single agent, resulting in high specialization. These results show that parallelizability plays a critical role in shaping the emergence of specialist or generalist behavior. 

\subsection{Experiment Design}
\label{experiment1design}

We evaluated teams of cooperative agents on two benchmark environments using JaxMARL \cite{flair2023jaxmarl}. In both environments, agents were trained independently using a recurrent actor–critic network and Independent PPO (IPPO) \cite{schulman2017proximal, lu2022discovered}. Full training details are provided in Appendix~\ref{training}. In \textbf{SMAC}, we systematically varied \emph{team size} ($N \in \{2,3,4,5,6 \}$) and \emph{task complexity}, defined as the number of enemy units per episode ($m \in \{2,3\}$). Each $(N, m)$ configuration was trained with 10 random seeds, resulting in 100 total runs. In \textbf{MPE}, we varied \emph{team size} ($N \in {2,3,5,10}$) and \emph{task complexity}, defined as the number of target landmarks ($m \in {2,3,5,10}$) which agents must cover simultaneously. Each $(N, m)$ configuration was trained with 10 random seeds, resulting in 160 total runs.

\subsection{Results}

\paragraph{Distribution of SI} Results are depicted in Figure~\ref{fig:smacmpe}. Experiment 1 revealed a clear relationship between task parallelizability and specialization across environments. In SMAC, agents always converged to generalist policies (mean SI $= 0.06$), with all trial-level mean SIs below $0.09$ and all raw SIs below $0.20$. Specialization decreased with increasing team size: the mean SI across agents was negatively correlated with $N$ (Pearson $r = -0.86$). This aligns with the prediction that SMAC has no task bottlenecks ($S(N, C) = N$), allowing  high parallelizability and low specialization. In contrast, MPE exhibited consistently high specialization across all conditions (mean SI $= 0.61$). Even in the best-performing seeds, SI remained above $0.5$, consistent with the environment’s strict spatial bottlenecks. Because each of the landmarks must be uniquely covered by one agent, the number of subtasks equals the number of agents and no two agents can complete the same subtask. Our parallelizability metric thus assigns $S = 1$ for MPE, predicting maximum specialization.

\paragraph{Model Fit}
To test the predicted relationship, we pooled data across environments and used logistic regression to predict whether agents would specialize based on the parallelizability score $S$. This model achieved perfect classification accuracy (1.0) across all trials. The relationship between $S$ and specialization was also strongly linear, with a Pearson correlation of $r = -0.84$ ($p = 5.9 \times 10^{-8}$).


\section{Experiment 2: Analysis of Mixed Bottlenecks in Overcooked-AI}
\label{experiment3}

\begin{figure*}[t]
     \centering
     \includegraphics[width = 0.9\linewidth]{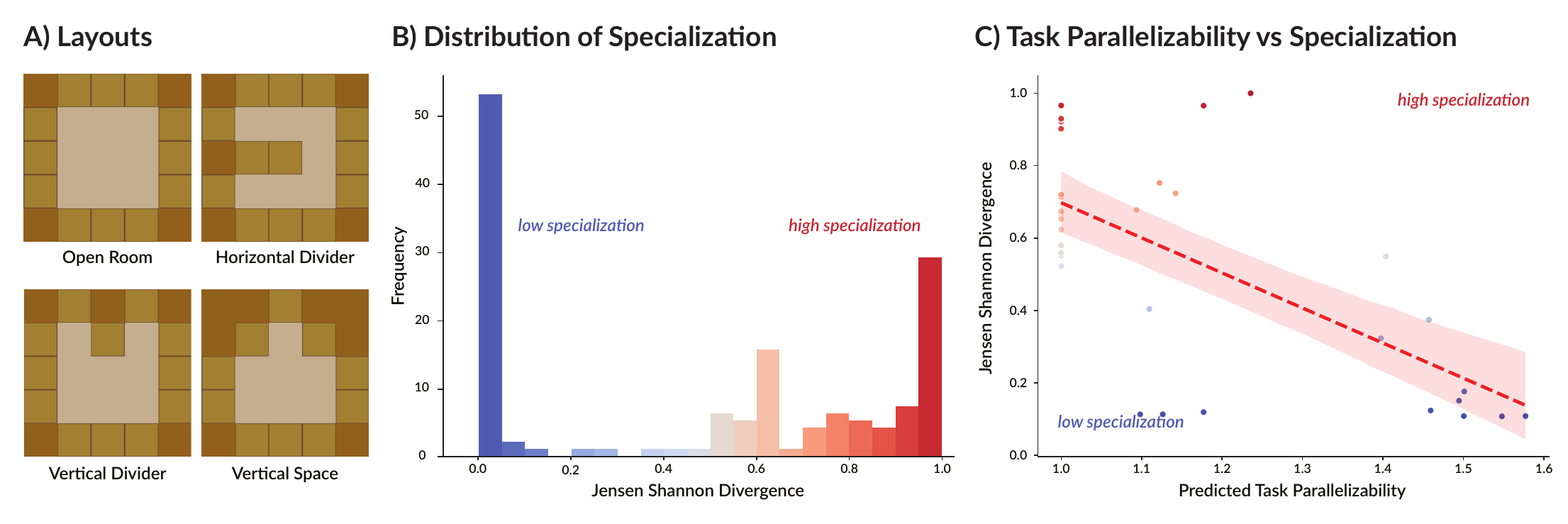}
    \caption{Experiment 2 Results. (A) Six $5 \times 5$ layouts with varying bottlenecks were paired with different recipes and pots for a total of 144 trials, 1440 runs. (B) Histogram of specialization (SI) in the best-performing seeds reveals a distinctly bimodal distribution, with most teams exhibiting low (generalist) or high (specialist) SI. (C) Scatter plot of task parallelizability ($S$) vs. observed SI. The negative correlation confirms that greater parallelizability corresponds to lower specialization. }
    \label{fig:results2}
\end{figure*}

Experiment 2 shows that both spatial and resource bottlenecks influence specialization, and that our parallelizability metric remains predictive under complex constraints. These findings suggest that parallelizability can be used to systematically design environments that promote or inhibit specialization, without the need to tailor training algorithms. When agents are given ample space and workstations, they develop generalist policies and learn the full scope of the task; in contrast, spatial or resource constraints lead to specialist policies. To test these predictions, we use Overcooked-AI, which features richer spatial topologies and resource constraints, supporting a wide range of strategies.

\subsection{Experiment Design}
\label{experiment2design}

We constructed six $5 \times 5$ layouts with varying \emph{spatial bottlenecks}: (1) open space with surrounding counters, (2) single vertical divider, (3) double vertical divider, (4) horizontal divider, (5) two dividers, and (6) long divider, illustrated in Figure~\ref{fig:results2}. We varied \emph{resource bottlenecks} (number of pots) while controlling for relative distances between workstations. Each layout featured four workstation permutations, totaling 48 configurations. Agents were trained to make one-, two-, or three-onion soup with 10 random seeds for a total of 1440 runs. For each trial, we trained two independent agents collaboratively via self-play using JaxMARL and iPPO \cite{schulman2017proximal, lu2022discovered, flair2023jaxmarl}. Training and compute details are in Appendix~\ref{training}. We selected the highest-performing seed based on joint reward during the roll-out. If multiple seeds reached the same reward, we averaged their SI. Trials were excluded if an agent failed to perform any successful action or the team scored a reward of 0. Four trials were excluded. 

\subsection{Results}

Results are shown in Figure \ref{fig:results2}. Our simulations revealed a clear bimodal distribution of specialization. There were 81 specialist teams with $SI \geq 0.5$ ($57.86\%$) and 59 generalist teams ($42.14\%$). Our task parallelizability model was generally successful at predicting specialization. The Pearson correlation between predicted task parallelizability and SI was $-0.67$ ($p = 8.94 \times 10^{-6}$), affirming the relationship between parallelizability and specialization. To validate the model's predictive power, we used logistic regression to classify high ($SI \geq 0.5$) or low specialization ($SI < 0.5$). The model achieved a test set accuracy of $90.90\%$, with $0.88$ precision ($F1 = 0.93$) for high specialization and $1.00$ precision ($F1=0.86$) for low. Details are in Appendix~\ref{experiment2_app}.

\section{Experiment 3: Exploratory Large-Scale Analysis in Overcooked-AI}
\label{experiment2}

In Experiment 2, we manipulated spatial and resource bottlenecks in controlled Overcooked environments, holding layout size and workstation counts (except for pots) constant. To more thoroughly examine how task parallelizability drives specialization, Experiment 3 expands the analysis by systematically varying layout size, bottleneck severity, and the number and arrangement of workstations.

\subsection{Experiment Design}
\label{experiment1design}

\begin{figure*}[t]
     \centering
     \includegraphics[width = \linewidth]{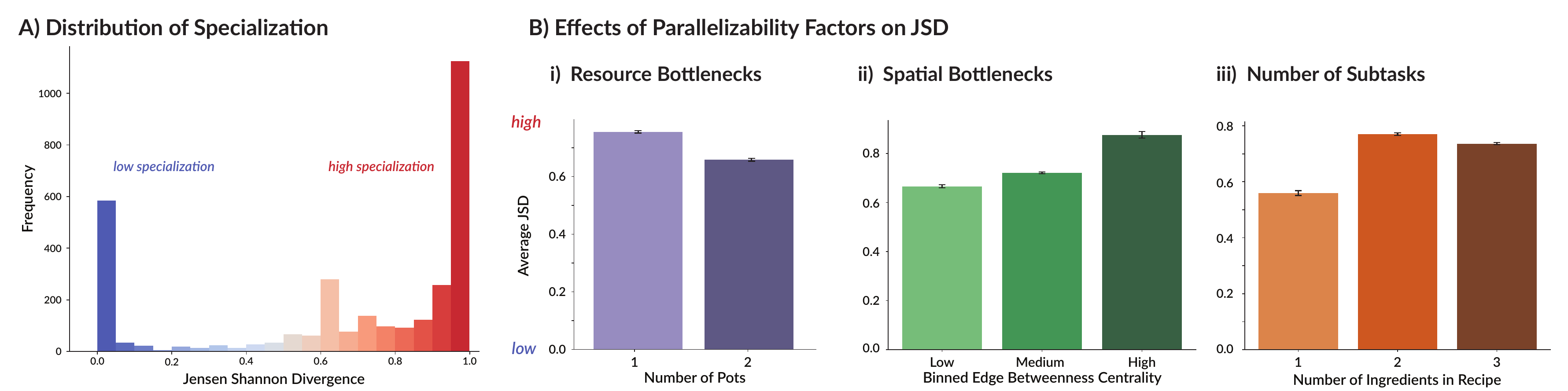}
    \caption{Experiment 3 Results. (A) Histogram of SI in best-performing seeds for 3,200 Overcooked configurations confirms bimodal distribution of generalists and specialists. (B) Variables influencing parallelizability show strong, significant effects on SI: specialization decreases greater concurrency (more pots, fewer spatial bottlenecks) and increases with subtasks. Error bars represent SEM. }
    \label{fig:results1}
\end{figure*}

We generated a test suite of 3,200 unique Overcooked environments for a large-scale exploratory analysis. We created 14 spatial configurations with unique arrangements of counters, including the original layouts from Overcooked-AI and a new set of layouts inspired by real-world kitchens \cite{pejic2019parametric}. A complete set of layouts is provided in Figure~\ref{fig:layouts}. To create diverse parallelizability constraints, we systematically varied the number of workstations (onions, tomatoes, pots, bowls, and serving stations). A total of 16 combinations included one or two onion piles, bowls, pots, serving stations, and zero or one tomatoes. Workstation positions were randomized three times per layout. Agents were trained to make (i) 1-onion (ii) 2-onion (iii) 3-onion (iv) 1-onion and 1-tomato and (v) 2-onion and 1-tomato soup. These variations produced a  3,200 unique environments. We used the  training procedures from Experiment 2, and again selected the highest-performing seed. 

\subsection{Results}
\label{experiment1results}

\textbf{Distribution of SI.} Results again revealed a distinctly bimodal distribution, yet more skewed towards specialization. Specialists ($SI \geq 0.5$) dominated $74.9\%$ of runs, with $48.9\%$ fully specialized ($SI > 0.9$). Generalists ($SI < 0.5$) appeared in $25.1\%$ of runs, including $22.3\%$ fully generalist ($SI < 0.1$). The remaining $28.8\%$ exhibited mixed specialist-generalist behaviors ($0.1 \leq SI \leq 0.9$).

\paragraph{Model Evaluation} Statistics can be found in Appendix~\ref{experiment1sigtests} and \ref{expt1_regression}, Tables \ref{tab:experiment1_layouts} and \ref{tab:quantities_SI}. For each of the variables presented in our predictive model, there was a significant effect in the expected direction on observed SI.  We found a significant effect of the number of pots, but not the total number of workstations; spatial bottlenecks; and recipe complexity (the number of ingredients required to make a soup) on emergent specialization. The average correlation $-0.486$ ($p=8.49 \times 10^{-28}$) revealed a moderate negative relationship between parallelizability and SI. A logistic regression classifying low vs.~high specialization ($SI \geq 0.5$) achieved $74.2\%$ accuracy, with $0.93$ precision ($F1 = 0.81$) for high specialization and $0.46$ precision ($F1 = 0.58$) for low specialization.

\paragraph{Effects of State Space Size} As illustrated, our model predictions deviated more from observed SI in this larger-scale analysis. Further investigation revealed a moderate correlation between grid size and SI ($r=0.383, p = 5.57 \times 10^{-17}$), even when controlling for bottlenecks (Figure~\ref{fig:gridsize}). In larger layouts, agents tended to converge to specialist policies, even with high predicted parallelizability.

\section{Experiment 4: Deviations from MARL Training Biases}

\label{experiment4design}
\begin{figure*}[t]
     \centering
     \includegraphics[width = 0.8\linewidth]{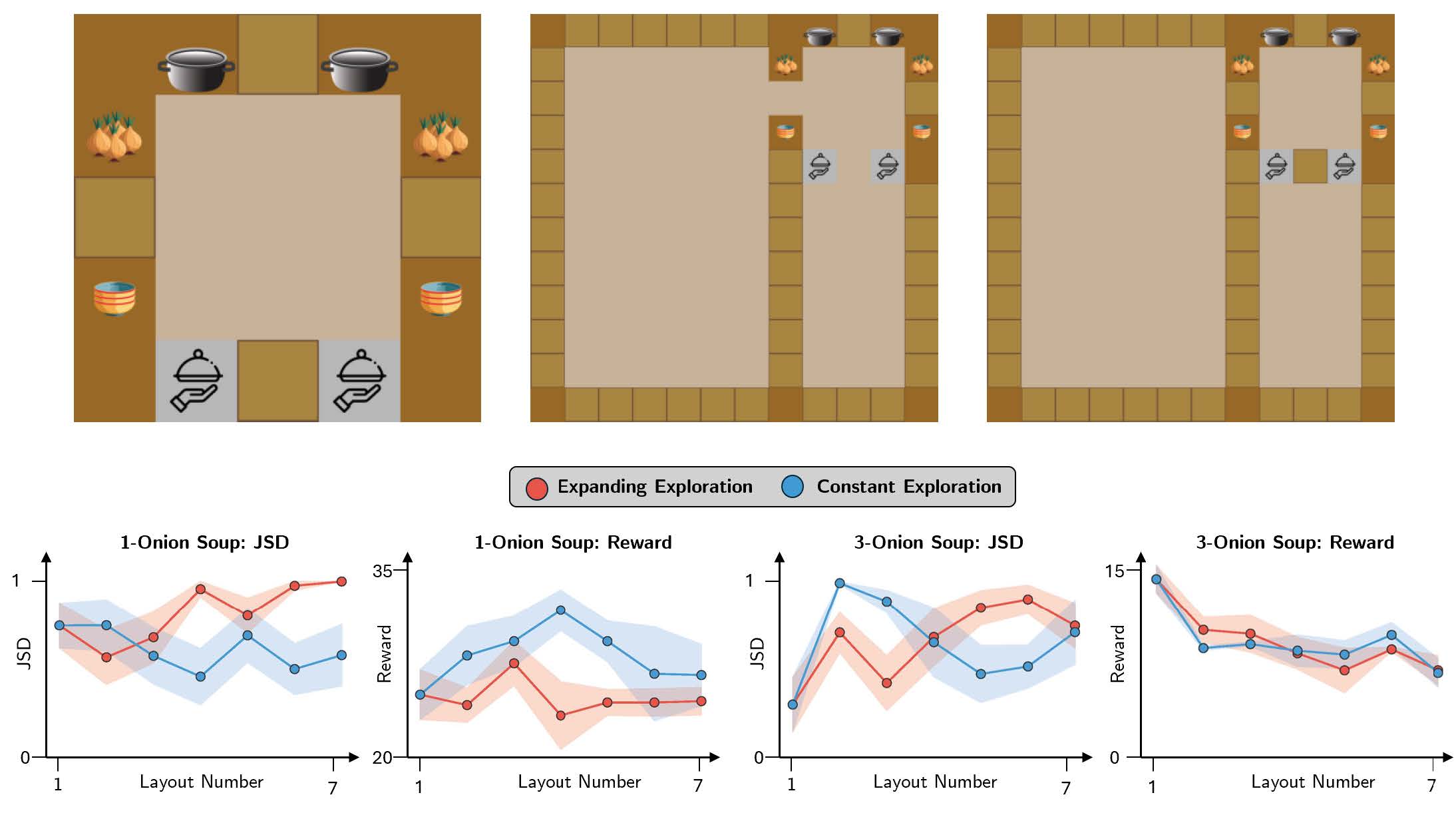}
    \caption{Experiment 4 Results. Experiment 2 and parallelizability predictions reveal that generalist policies are most efficient in each layout. However, specialization significantly increases with layout size, also leading to lower reward for more complex tasks. }
    \label{fig:experiment4}
\end{figure*}

In Experiment 4, we test whether deviations from our predictions in Experiment 3 stem from limits of our parallelizability framework or from increased exploration demands in larger state spaces. We find that as state space grows, despite a fixed grid size that is needed to complete the task, agents adopt more specialized strategies and achieve lower rewards with larger recipes. This suggests that in complex environments, specialization serves as a tractable alternative, enabling agents to focus on restricted subspaces when optimal joint policy learning is challenging.


\subsection{Experiment Design}
We re-used the first four $5 \times 5$ layouts with two pots from Experiment 2 (depicted in Figure~\ref{fig:experiment4}) and systematically varied seven layout sizes and two conditions. In the constant condition, we increased the overall grid size but held the amount of floor space that agents could access constant. In the non-constant condition, we increased the overall grid size and allowed agents to access it. Critically, Experiment 2 confirmed that the optimal behavior for these layouts were generalists; although the size of the state space differs, the path lengths required to actually complete the tasks remained the same. In total, there were 280 runs (7 sizes, 2 conditions, 2 recipes, and 10 random seeds).

\subsection{Results}

\paragraph{Distribution of SI} For the one-onion recipe, agents consistently converged to fully specialist policies in three of the layouts ($100\%$ with a single divider, $81.3\%$ with a double divider, and $81.3\%$ with no-divider), whereas only $46.3\%$ of runs in the counter circuit resulted in SI $\geq 0.5$. For the three-onion soup recipe, agents also consistently converged to specialist policies ($93.8\%$ with a single divider, $87.5\%$ with a double divider, $87.5\%$ with no-divider, and $93.8\%$ in the counter circuit). 

\paragraph{Interaction Between State Space Size and Task Complexity} The effect of layout size varied across the two recipes, indicating that both state space size and the number of tasks create a more difficult learning problem. With one-onion soup, SI increased significantly with layout size ($r = 0.7924, p = 0.03$). While there was a negative relationship between throughput and improper specialization, it was not statistically significant ($r = -0.2512, p = 0.5869$). In contrast, for the three-onion soup task, SI increased significantly with layout size ($r = 0.7549, p = 0.05$) and throughput significantly decreased ($r = -0.8556, p = 0.0140$), confirming that deviations from generalist strategies when possible lead to worse team performance (Proposition 4).  

\section{Related Work}
The study of specialization in multi-agent systems spans multiple disciplines. Division of labor leading to functional specialization has been widely studied in biological and social domains, including cells \cite{bell1997size, rueffler2012evolution}, insect colonies \cite{ratnieks1999task, fjerdingstad2006evolution}, and human societies \cite{kitcher1990division, kreider2022resource, goldstone2024emergence, velez2024rise}. Inspired by these insights, research in MARL and robotics has sought to design algorithms that actively encourage specialization by promoting policy diversity \cite{murciano1997specialization, padgham2002prometheus, zhu2008role, charakorn2020investigating, wang2020roma, wang2020rode, li2021celebrating, bettini2023heterogeneous, bettini2024controlling} or by organizing agents into hierarchies \cite{zhang2010self, ahilan2019feudal}. Similarly, new work on language-agent collaboration often assumes that specialization is necessary, prompting agents to adopt distinct roles \cite{wang2023unleashing, suzgun2024meta, swanson2024virtual, liu2024dynamic, juang2024breaking}. However, while biological and social systems provide compelling examples, much of this work assumes that specialization is beneficial without systematically identifying when and why generalists might outperform specialists. For example, one benefit of generalists is their flexibility, which allows them to outperform rigidly specified roles under conditions of high environmental variability or individual failure \cite{staps2022being, ben2023cultural}.

A useful contrast comes from distributed systems, which are explicitly engineered rather than emergent. In these systems, heterogeneity and parallelism are carefully designed to optimize performance \cite{moncrieff1996heterogeneous, almasi1994highly, pancake1996parallelism} and minimize contention \cite{park2001deadlock}. This structured approach provides a useful reference point for understanding when emergent specialization is not just incidental, but optimal. In our work, we propose a novel theoretical framework for predicting the utility of specialization based on task efficiency, considering environmental bottlenecks, resource constraints, and task complexity. Prior work in RL has shown that spatial bottlenecks such as doorways can shape subgoals \cite{menache2002q}, while research in human-robot collaboration and MARL has demonstrated that environmental design influences emergent behavior \cite{fontaine2021importance, mckee2022quantifying}. Building on these insights, we formalize a predictive model for when generalist versus specialist strategies should emerge.

\section{Discussion}
\label{discussion}

Specialization, or the differentiation of agents' policies and roles, is often assumed to improve team performance. However, principles from distributed systems offer an alternative hypothesis. Specifically, generalist agents may outperform specialists when concurrent execution of the task would improve speed and efficiency. In this work, we outlined a theoretical framework for predicting generalist versus specialist efficiency in cooperative multi-agent tasks inspired by a variant of Amdahl's Law. We then validated this model across core MARL environments (SMAC and MPE), as well as three in-depth Overcooked-AI experiments. Together, these findings suggest that generalists are constrained by their environment and task parallelizability, making it possible to design layouts which either encourage or inhibit role emergence regardless of whether the training algorithm explicitly incentivizes diversity. Our framework is highly scalable, providing principled predictions for emergent behavior given various team sizes, environments, and task constraints.  More broadly, this framework motivates the use of distributed systems in analyzing trade-offs in multi-agent systems. 

This work has several limitations. First, our model does not account for switching behavior (e.g., generalists who alternate roles) or cases where specialist and generalist policies perform equally well, reflecting an implicit assumption from Amdahl’s Law that subtasks are executed sequentially without overlap. Second, while our model provides a theoretical upper bound on the benefits of generalists, it omits coordination overhead, which may make the empirical relationship non-monotonic. Third, our model assumes agents are initially homogeneous, yet specialization can also arise when agents have heterogeneous skills that make them more efficient at certain tasks \cite{moncrieff1996heterogeneous, ginting2022capability}. Finally, Experiment 4 suggests that specialization is a locally optimal solution that agents use when exploration is costly, i.e. in larger state spaces. Clarifying this trade-off can provide insight into how specialization may improve sample efficiency, as well as how to decrease coordination overhead between agents. 


\bibliography{references}

\begin{thebibliography}{66}
\providecommand{\natexlab}[1]{#1}
\providecommand{\url}[1]{\texttt{#1}}
\expandafter\ifx\csname urlstyle\endcsname\relax
  \providecommand{\doi}[1]{doi: #1}\else
  \providecommand{\doi}{doi: \begingroup \urlstyle{rm}\Url}\fi

\bibitem[Canese et~al.(2021)Canese, Cardarilli, Di~Nunzio, Fazzolari, Giardino, Re, and Span{\`o}]{canese2021multi}
Lorenzo Canese, Gian~Carlo Cardarilli, Luca Di~Nunzio, Rocco Fazzolari, Daniele Giardino, Marco Re, and Sergio Span{\`o}.
\newblock Multi-agent reinforcement learning: A review of challenges and applications.
\newblock \emph{Applied Sciences}, 11\penalty0 (11):\penalty0 4948, 2021.

\bibitem[Perolat et~al.(2017)Perolat, Leibo, Zambaldi, Beattie, Tuyls, and Graepel]{perolat2017multi}
Julien Perolat, Joel~Z Leibo, Vinicius Zambaldi, Charles Beattie, Karl Tuyls, and Thore Graepel.
\newblock A multi-agent reinforcement learning model of common-pool resource appropriation.
\newblock \emph{Advances in Neural Information Processing Systems}, 30, 2017.

\bibitem[Park et~al.(2023)Park, O’Brien, Cai, Morris, Liang, and Bernstein]{park2023generative}
Joon~Sung Park, Joseph~C O’Brien, Carrie~J Cai, Meredith~Ringel Morris, Percy Liang, and Michael~S Bernstein.
\newblock Generative agents: Interactive simulacra of human behavior. arxiv.
\newblock \emph{arXiv preprint ArXiv:2304.03442}, 2023.

\bibitem[Padgham and Winikoff(2002)]{padgham2002prometheus}
Lin Padgham and Michael Winikoff.
\newblock Prometheus: A methodology for developing intelligent agents.
\newblock In \emph{{Proceedings of the First International Joint Conference on Autonomous Agents and Multiagent Systems: Part 1}}, pages 37--38, 2002.

\bibitem[Zhu and Zhou(2008)]{zhu2008role}
Haibin Zhu and MengChu Zhou.
\newblock Role-based multi-agent systems.
\newblock In \emph{Personalized Information Retrieval and Access: Concepts, Methods and Practices}, pages 254--285. IGI Global, 2008.

\bibitem[Charakorn et~al.(2020)Charakorn, Manoonpong, and Dilokthanakul]{charakorn2020investigating}
Rujikorn Charakorn, Poramate Manoonpong, and Nat Dilokthanakul.
\newblock Investigating partner diversification methods in cooperative multi-agent deep reinforcement learning.
\newblock In \emph{{International Conference on Neural Information Processing}}, pages 395--402, 2020.

\bibitem[Wang et~al.(2020{\natexlab{a}})Wang, Dong, Lesser, and Zhang]{wang2020roma}
Tonghan Wang, Heng Dong, Victor Lesser, and Chongjie Zhang.
\newblock Roma: Multi-agent reinforcement learning with emergent roles.
\newblock \emph{arXiv preprint arXiv:2003.08039}, 2020{\natexlab{a}}.

\bibitem[Wang et~al.(2020{\natexlab{b}})Wang, Gupta, Mahajan, Peng, Whiteson, and Zhang]{wang2020rode}
Tonghan Wang, Tarun Gupta, Anuj Mahajan, Bei Peng, Shimon Whiteson, and Chongjie Zhang.
\newblock Rode: Learning roles to decompose multi-agent tasks.
\newblock \emph{arXiv preprint arXiv:2010.01523}, 2020{\natexlab{b}}.

\bibitem[Li et~al.(2021)Li, Wang, Wu, Zhao, Yang, and Zhang]{li2021celebrating}
Chenghao Li, Tonghan Wang, Chengjie Wu, Qianchuan Zhao, Jun Yang, and Chongjie Zhang.
\newblock Celebrating diversity in shared multi-agent reinforcement learning.
\newblock \emph{Advances in Neural Information Processing Systems}, 34:\penalty0 3991--4002, 2021.

\bibitem[Bettini et~al.(2023)Bettini, Shankar, and Prorok]{bettini2023heterogeneous}
Matteo Bettini, Ajay Shankar, and Amanda Prorok.
\newblock Heterogeneous multi-robot reinforcement learning.
\newblock \emph{arXiv preprint arXiv:2301.07137}, 2023.

\bibitem[Juang et~al.(2024)Juang, Cao, Zhou, Liu, Zhang, and Liu]{juang2024breaking}
Stefan Juang, Hugh Cao, Arielle Zhou, Ruochen Liu, Nevin~L Zhang, and Elvis Liu.
\newblock Breaking the mold: The challenge of large scale {MARL} specialization.
\newblock \emph{arXiv preprint arXiv:2410.02128}, 2024.

\bibitem[Bettini et~al.(2024)Bettini, Kortvelesy, and Prorok]{bettini2024controlling}
Matteo Bettini, Ryan Kortvelesy, and Amanda Prorok.
\newblock Controlling behavioral diversity in multi-agent reinforcement learning.
\newblock \emph{arXiv preprint arXiv:2405.15054}, 2024.

\bibitem[Swanson et~al.(2024)Swanson, Wu, Bulaong, Pak, and Zou]{swanson2024virtual}
Kyle Swanson, Wesley Wu, Nash~L Bulaong, John~E Pak, and James Zou.
\newblock The virtual lab: {AI} agents design new {Sars-Cov-2} nanobodies with experimental validation.
\newblock \emph{bioRxiv}, pages 2024--11, 2024.

\bibitem[Goldstone et~al.(2024)Goldstone, Andrade-Lotero, Hawkins, and Roberts]{goldstone2024emergence}
Robert~L Goldstone, Edgar~J Andrade-Lotero, Robert~D Hawkins, and Michael~E Roberts.
\newblock The emergence of specialized roles within groups.
\newblock \emph{{Topics in Cognitive Science}}, 16\penalty0 (2):\penalty0 257--281, 2024.

\bibitem[Griffiths(2020)]{griffiths2020understanding}
Thomas~L Griffiths.
\newblock Understanding human intelligence through human limitations.
\newblock \emph{Trends in Cognitive Sciences}, 24\penalty0 (11):\penalty0 873--883, 2020.

\bibitem[Ben-Oren et~al.(2023)Ben-Oren, Kolodny, and Creanza]{ben2023cultural}
Yotam Ben-Oren, Oren Kolodny, and Nicole Creanza.
\newblock Cultural specialization as a double-edged sword: division into specialized guilds might promote cultural complexity at the cost of higher susceptibility to cultural loss.
\newblock \emph{Philosophical Transactions of the Royal Society B}, 378\penalty0 (1872):\penalty0 20210418, 2023.

\bibitem[V{\'e}lez et~al.(2024)V{\'e}lez, Wu, Gershman, and Schulz]{velez2024rise}
Natalia V{\'e}lez, Charley~M Wu, Samuel~J Gershman, and Eric Schulz.
\newblock The rise and fall of technological development in virtual communities.
\newblock \emph{PsyArXiv}, 2024.

\bibitem[Shoham and Leyton-Brown(2008)]{shoham2008multiagent}
Yoav Shoham and Kevin Leyton-Brown.
\newblock \emph{Multiagent systems: Algorithmic, game-theoretic, and logical foundations}.
\newblock Cambridge University Press, 2008.

\bibitem[Wu et~al.(2021)Wu, Wang, Evans, Tenenbaum, Parkes, and Kleiman-Weiner]{wu2021too}
Sarah~A Wu, Rose~E Wang, James~A Evans, Joshua~B Tenenbaum, David~C Parkes, and Max Kleiman-Weiner.
\newblock Too many cooks: Bayesian inference for coordinating multi-agent collaboration.
\newblock \emph{{Topics in Cognitive Science}}, 13\penalty0 (2):\penalty0 414--432, 2021.

\bibitem[Almasi and Gottlieb(1994)]{almasi1994highly}
George~S Almasi and Allan Gottlieb.
\newblock \emph{Highly parallel computing}.
\newblock Benjamin-Cummings Publishing Co., 1994.

\bibitem[McCool et~al.(2012)McCool, Reinders, and Robison]{mccool2012structured}
Michael McCool, James Reinders, and Arch Robison.
\newblock \emph{Structured parallel programming: {Patterns} for efficient computation}.
\newblock Elsevier, 2012.

\bibitem[Amdahl(1967)]{amdahl1967validity}
Gene~M Amdahl.
\newblock Validity of the single processor approach to achieving large scale computing capabilities.
\newblock In \emph{{Proceedings of the {April} 18-20, 1967, Spring Joint Computer Conference}}, pages 483--485, 1967.

\bibitem[Hill and Marty(2008)]{hill2008amdahl}
Mark~D Hill and Michael~R Marty.
\newblock {Amdahl's} law in the multicore era.
\newblock \emph{{Computer}}, 41\penalty0 (7):\penalty0 33--38, 2008.

\bibitem[Cassidy and Andreou(2011)]{cassidy2011beyond}
Andrew~S Cassidy and Andreas~G Andreou.
\newblock Beyond {Amdahl's} law: An objective function that links multiprocessor performance gains to delay and energy.
\newblock \emph{{IEEE Transactions on Computers}}, 61\penalty0 (8):\penalty0 1110--1126, 2011.

\bibitem[Albrecht et~al.(2024)Albrecht, Christianos, and Sch{\"a}fer]{albrecht2024multi}
Stefano~V Albrecht, Filippos Christianos, and Lukas Sch{\"a}fer.
\newblock \emph{Multi-agent reinforcement learning: Foundations and modern approaches}.
\newblock MIT Press, 2024.

\bibitem[Schulman et~al.(2015)Schulman, Levine, Abbeel, Jordan, and Moritz]{schulman2015trust}
John Schulman, Sergey Levine, Pieter Abbeel, Michael Jordan, and Philipp Moritz.
\newblock Trust region policy optimization.
\newblock In \emph{International conference on machine learning}, pages 1889--1897. PMLR, 2015.

\bibitem[Endres and Schindelin(2003)]{endres2003new}
Dominik~Maria Endres and Johannes~E Schindelin.
\newblock A new metric for probability distributions.
\newblock \emph{{IEEE Transactions on Information Theory}}, 49\penalty0 (7):\penalty0 1858--1860, 2003.

\bibitem[Fuglede and Topsoe(2004)]{fuglede2004jensen}
Bent Fuglede and Flemming Topsoe.
\newblock {Jensen-Shannon} divergence and {Hilbert} space embedding.
\newblock In \emph{{International Symposium on Information Theory}}, page~31, 2004.

\bibitem[Zhuravlev et~al.(2010)Zhuravlev, Blagodurov, and Fedorova]{zhuravlev2010addressing}
Sergey Zhuravlev, Sergey Blagodurov, and Alexandra Fedorova.
\newblock Addressing shared resource contention in multicore processors via scheduling.
\newblock \emph{ACM Sigplan Notices}, 45\penalty0 (3):\penalty0 129--142, 2010.

\bibitem[Dublish et~al.(2017)Dublish, Nagarajan, and Topham]{dublish2017evaluating}
Saumay Dublish, Vijay Nagarajan, and Nigel Topham.
\newblock Evaluating and mitigating bandwidth bottlenecks across the memory hierarchy in gpus.
\newblock In \emph{{IEEE International Symposium on Performance Analysis of Systems and Software (ISPASS)}}, pages 239--248, 2017.

\bibitem[{\v{Z}}ivi{\v{c}}njak et~al.(2022){\v{Z}}ivi{\v{c}}njak, Rogi{\'c}, and Bajor]{vzivivcnjak2022case}
Margareta {\v{Z}}ivi{\v{c}}njak, Kristijan Rogi{\'c}, and Ivona Bajor.
\newblock Case-study analysis of warehouse process optimization.
\newblock \emph{Transportation Research Procedia}, 64:\penalty0 215--223, 2022.

\bibitem[Zhang and Gao(2020)]{zhang2020will}
Tianshu Zhang and Kun Gao.
\newblock Will autonomous vehicles improve traffic efficiency and safety in urban road bottlenecks?
\newblock In \emph{International Conference on Intelligent Transportation Engineering (ICITE)}, pages 366--370, 2020.

\bibitem[Wang et~al.(2016)Wang, Wan, Zhang, Li, and Zhang]{wang2016towards}
Shiyong Wang, Jiafu Wan, Daqiang Zhang, Di~Li, and Chunhua Zhang.
\newblock Towards smart factory for industry 4.0: A self-organized multi-agent system with big data based feedback and coordination.
\newblock \emph{Computer Networks}, 101:\penalty0 158--168, 2016.

\bibitem[Mieczkowski et~al.(2024)Mieczkowski, Turner, V{\'e}lez, and Griffiths]{mieczkowskipeople}
Elizabeth~A Mieczkowski, Cameron~Rouse Turner, Natalia V{\'e}lez, and Thomas~L Griffiths.
\newblock People evaluate idle collaborators based on their impact on task efficiency.
\newblock \emph{PsyArXiv}, 2024.

\bibitem[Solway et~al.(2014)Solway, Diuk, C{\'o}rdova, Yee, Barto, Niv, and Botvinick]{solway2014optimal}
Alec Solway, Carlos Diuk, Natalia C{\'o}rdova, Debbie Yee, Andrew~G Barto, Yael Niv, and Matthew~M Botvinick.
\newblock Optimal behavioral hierarchy.
\newblock \emph{PLoS computational biology}, 10\penalty0 (8):\penalty0 e1003779, 2014.

\bibitem[Amato et~al.(2019)Amato, Konidaris, Kaelbling, and How]{amato2019modeling}
Christopher Amato, George Konidaris, Leslie~P Kaelbling, and Jonathan~P How.
\newblock Modeling and planning with macro-actions in decentralized pomdps.
\newblock \emph{Journal of Artificial Intelligence Research}, 64:\penalty0 817--859, 2019.

\bibitem[De~Witt et~al.(2020)De~Witt, Gupta, Makoviichuk, Makoviychuk, Torr, Sun, and Whiteson]{de2020independent}
Christian~Schroeder De~Witt, Tarun Gupta, Denys Makoviichuk, Viktor Makoviychuk, Philip~HS Torr, Mingfei Sun, and Shimon Whiteson.
\newblock Is independent learning all you need in the starcraft multi-agent challenge?
\newblock \emph{arXiv preprint arXiv:2011.09533}, 2020.

\bibitem[Lowe et~al.(2017)Lowe, Wu, Tamar, Harb, Pieter~Abbeel, and Mordatch]{lowe2017multi}
Ryan Lowe, Yi~I Wu, Aviv Tamar, Jean Harb, OpenAI Pieter~Abbeel, and Igor Mordatch.
\newblock Multi-agent actor-critic for mixed cooperative-competitive environments.
\newblock \emph{Advances in neural information processing systems}, 30, 2017.

\bibitem[Carroll et~al.(2019)Carroll, Shah, Ho, Griffiths, Seshia, Abbeel, and Dragan]{carroll2019utility}
Micah Carroll, Rohin Shah, Mark~K Ho, Tom Griffiths, Sanjit Seshia, Pieter Abbeel, and Anca Dragan.
\newblock On the utility of learning about humans for human-{AI} coordination.
\newblock \emph{{Advances in Neural Information Processing Systems}}, 32, 2019.

\bibitem[Barrat et~al.(2004)Barrat, Barthelemy, Pastor-Satorras, and Vespignani]{barrat2004architecture}
Alain Barrat, Marc Barthelemy, Romualdo Pastor-Satorras, and Alessandro Vespignani.
\newblock The architecture of complex weighted networks.
\newblock \emph{{Proceedings of the National Academy of Sciences}}, 101\penalty0 (11):\penalty0 3747--3752, 2004.

\bibitem[Rutherford et~al.(2023)Rutherford, Ellis, Gallici, Cook, Lupu, Ingvarsson, Willi, Khan, de~Witt, Souly, Bandyopadhyay, Samvelyan, Jiang, Lange, Whiteson, Lacerda, Hawes, Rocktaschel, Lu, and Foerster]{flair2023jaxmarl}
Alexander Rutherford, Benjamin Ellis, Matteo Gallici, Jonathan Cook, Andrei Lupu, Gardar Ingvarsson, Timon Willi, Akbir Khan, Christian~Schroeder de~Witt, Alexandra Souly, Saptarashmi Bandyopadhyay, Mikayel Samvelyan, Minqi Jiang, Robert~Tjarko Lange, Shimon Whiteson, Bruno Lacerda, Nick Hawes, Tim Rocktaschel, Chris Lu, and Jakob~Nicolaus Foerster.
\newblock Jaxmarl: Multi-agent rl environments in jax.
\newblock \emph{arXiv preprint arXiv:2311.10090}, 2023.

\bibitem[Schulman et~al.(2017)Schulman, Wolski, Dhariwal, Radford, and Klimov]{schulman2017proximal}
John Schulman, Filip Wolski, Prafulla Dhariwal, Alec Radford, and Oleg Klimov.
\newblock Proximal policy optimization algorithms.
\newblock \emph{arXiv}, 2017.

\bibitem[Lu et~al.(2022)Lu, Kuba, Letcher, Metz, de~Witt, and Foerster]{lu2022discovered}
Chris Lu, Jakub~Grudzien Kuba, Alistair Letcher, Luke Metz, Christian~Schroeder de~Witt, and Jakob Foerster.
\newblock Discovered policy optimisation.
\newblock \emph{{Advances in Neural Information Processing Systems}}, 2022.

\bibitem[Pejic et~al.(2019)Pejic, Jovanovic, Marinkovic, Stojakovic, and Krasic]{pejic2019parametric}
Petar Pejic, Dimitrije Jovanovic, Jovan Marinkovic, Vesna Stojakovic, and Sonja Krasic.
\newblock Parametric {3D modeling of I-shape} kitchen.
\newblock \emph{{Journal of Industrial Design and Engineering Graphics}}, 14\penalty0 (1):\penalty0 155--158, 2019.

\bibitem[Bell and Mooers(1997)]{bell1997size}
Graham Bell and Arne~O Mooers.
\newblock Size and complexity among multicellular organisms.
\newblock \emph{Biological Journal of the Linnean Society}, 60\penalty0 (3):\penalty0 345--363, 1997.

\bibitem[Rueffler et~al.(2012)Rueffler, Hermisson, and Wagner]{rueffler2012evolution}
Claus Rueffler, Joachim Hermisson, and G{\"u}nter~P Wagner.
\newblock Evolution of functional specialization and division of labor.
\newblock \emph{{Proceedings of the National Academy of Sciences}}, 109\penalty0 (6):\penalty0 E326--E335, 2012.

\bibitem[Ratnieks and Anderson(1999)]{ratnieks1999task}
Francis~LW Ratnieks and Carl Anderson.
\newblock Task partitioning in insect societies.
\newblock \emph{Insectes Sociaux}, 46:\penalty0 95--108, 1999.

\bibitem[Fjerdingstad and Crozier(2006)]{fjerdingstad2006evolution}
Else~J Fjerdingstad and Ross~H Crozier.
\newblock The evolution of worker caste diversity in social insects.
\newblock \emph{The American Naturalist}, 167\penalty0 (3):\penalty0 390--400, 2006.

\bibitem[Kitcher(1990)]{kitcher1990division}
Philip Kitcher.
\newblock The division of cognitive labor.
\newblock \emph{The journal of philosophy}, 87\penalty0 (1):\penalty0 5--22, 1990.

\bibitem[Kreider et~al.(2022)Kreider, Janzen, Bernadou, Elsner, Kramer, and Weissing]{kreider2022resource}
Jan~J Kreider, Thijs Janzen, Abel Bernadou, Daniel Elsner, Boris~H Kramer, and Franz~J Weissing.
\newblock Resource sharing is sufficient for the emergence of division of labour.
\newblock \emph{Nature Communications}, 13\penalty0 (1):\penalty0 7232, 2022.

\bibitem[Murciano et~al.(1997)Murciano, del R.~Mill{\'a}n, and Zamora]{murciano1997specialization}
Antonio Murciano, Jos{\'e} del R.~Mill{\'a}n, and Javier Zamora.
\newblock Specialization in multi-agent systems through learning.
\newblock \emph{Biological Cybernetics}, 76\penalty0 (5):\penalty0 375--382, 1997.

\bibitem[Zhang et~al.(2010)Zhang, Lesser, and Abdallah]{zhang2010self}
Chongjie Zhang, Victor~R Lesser, and Sherief Abdallah.
\newblock Self-organization for coordinating decentralized reinforcement learning.
\newblock In \emph{{{Proceedings of the International Joint Conference on Autonomous Agents and Multiagent Systems}}}, volume~10, pages 739--746, 2010.

\bibitem[Ahilan and Dayan(2019)]{ahilan2019feudal}
Sanjeevan Ahilan and Peter Dayan.
\newblock Feudal multi-agent hierarchies for cooperative reinforcement learning.
\newblock \emph{arXiv preprint arXiv:1901.08492}, 2019.

\bibitem[Wang et~al.(2023)Wang, Mao, Wu, Ge, Wei, and Ji]{wang2023unleashing}
Zhenhailong Wang, Shaoguang Mao, Wenshan Wu, Tao Ge, Furu Wei, and Heng Ji.
\newblock Unleashing the emergent cognitive synergy in large language models: A task-solving agent through multi-persona self-collaboration.
\newblock \emph{arXiv preprint arXiv:2307.05300}, 2023.

\bibitem[Suzgun and Kalai(2024)]{suzgun2024meta}
Mirac Suzgun and Adam~Tauman Kalai.
\newblock Meta-prompting: Enhancing language models with task-agnostic scaffolding.
\newblock \emph{arXiv preprint arXiv:2401.12954}, 2024.

\bibitem[Liu et~al.(2024)Liu, Zhang, Li, Liu, and Yang]{liu2024dynamic}
Zijun Liu, Yanzhe Zhang, Peng Li, Yang Liu, and Diyi Yang.
\newblock A dynamic {LLM}-powered agent network for task-oriented agent collaboration.
\newblock In \emph{{First Conference on Language Modeling}}, 2024.

\bibitem[Staps and Tarnita(2022)]{staps2022being}
Merlijn Staps and Corina~E Tarnita.
\newblock When being flexible matters: Ecological underpinnings for the evolution of collective flexibility and task allocation.
\newblock \emph{{Proceedings of the National Academy of Sciences}}, 119\penalty0 (18):\penalty0 e2116066119, 2022.

\bibitem[Moncrieff et~al.(1996)Moncrieff, Overill, and Wilson]{moncrieff1996heterogeneous}
David Moncrieff, Richard~E. Overill, and Stephen Wilson.
\newblock Heterogeneous computing machines and {Amdahl's Law}.
\newblock \emph{Parallel Computing}, 22\penalty0 (3):\penalty0 407--413, 1996.

\bibitem[Pancake(1996)]{pancake1996parallelism}
C~Pancake.
\newblock Is parallelism for you?
\newblock \emph{IEEE Computational Science and Engineering}, 3\penalty0 (2):\penalty0 18--37, 1996.

\bibitem[Park and Reveliotis(2001)]{park2001deadlock}
Jonghun Park and Spyros~A Reveliotis.
\newblock Deadlock avoidance in sequential resource allocation systems with multiple resource acquisitions and flexible routings.
\newblock \emph{IEEE Transactions on Automatic Control}, 46\penalty0 (10):\penalty0 1572--1583, 2001.

\bibitem[Menache et~al.(2002)Menache, Mannor, and Shimkin]{menache2002q}
Ishai Menache, Shie Mannor, and Nahum Shimkin.
\newblock Q-cut—dynamic discovery of sub-goals in reinforcement learning.
\newblock In \emph{{13th European Conference on Machine Learning}}, pages 295--306, 2002.

\bibitem[Fontaine et~al.(2021)Fontaine, Hsu, Zhang, Tjanaka, and Nikolaidis]{fontaine2021importance}
Matthew~C Fontaine, Ya-Chuan Hsu, Yulun Zhang, Bryon Tjanaka, and Stefanos Nikolaidis.
\newblock On the importance of environments in human-robot coordination.
\newblock \emph{arXiv preprint arXiv:2106.10853}, 2021.

\bibitem[McKee et~al.(2022)McKee, Leibo, Beattie, and Everett]{mckee2022quantifying}
Kevin~R McKee, Joel~Z Leibo, Charlie Beattie, and Richard Everett.
\newblock Quantifying the effects of environment and population diversity in multi-agent reinforcement learning.
\newblock \emph{Autonomous Agents and Multi-Agent Systems}, 36\penalty0 (1):\penalty0 21, 2022.

\bibitem[Ginting et~al.(2022)Ginting, Otsu, Kochenderfer, and Agha-Mohammadi]{ginting2022capability}
Muhammad~Fadhil Ginting, Kyohei Otsu, Mykel~J Kochenderfer, and Ali-Akbar Agha-Mohammadi.
\newblock Capability-aware task allocation and team formation analysis for cooperative exploration of complex environments.
\newblock In \emph{{International Conference on Intelligent Robots and Systems}}, pages 7145--7152, 2022.

\bibitem[Haarnoja et~al.(2018)Haarnoja, Zhou, Abbeel, and Levine]{haarnoja2018soft}
Tuomas Haarnoja, Aurick Zhou, Pieter Abbeel, and Sergey Levine.
\newblock Soft actor-critic: Off-policy maximum entropy deep reinforcement learning with a stochastic actor.
\newblock In \emph{{International Conference on Machine Learning}}, pages 1861--1870, 2018.

\bibitem[Zhang(2018)]{zhang2018improved}
Zijun Zhang.
\newblock Improved {Adam} optimizer for deep neural networks.
\newblock In \emph{{International Symposium on Quality of Service}}, pages 1--2, 2018.

\end{thebibliography}
\bibliographystyle{unsrtnat}

\newpage
\appendix

\section{State-Visitation and Action Distributions}
\label{stateactionoccupancy}

In order to measure the differences between agents' policies, we must quantify how different their distributions of states and actions are. Each agent follows a state-visitation distribution that describes how frequently they occupy different states under their policy \cite{schulman2015trust} given by:
\begin{equation}
    d^\pi(s) = (1 - \gamma) \sum_{t=0}^{\infty} \gamma^t P(s_t = s | \pi)
\end{equation}
The corresponding state-action visitation distribution is:
\begin{equation}
    d^\pi(s, a) = d^\pi(s) \pi(a | s),
\end{equation}
which represents the frequency with which an agent selects action $a$ in state $s$ under policy $\pi$. For each agent $i$, we define their effective action distribution using their visitation distribution:
\begin{equation}
P_i(a) = \sum_{s \in \mathcal{S}} d^\pi(s, a)
\end{equation}
This distribution captures both what actions an agent tends to take and how often they visit states where those actions are available.

\section{Original Form of Amdahl's Law}
\label{originalamdahl}
Let $f$ represent the proportion of a task that can be parallelized and accelerated by a factor of $s$, while $1-f$ denotes the portion that must be executed serially. 
The theoretical speedup $S$ achievable when distributing a task across multiple processors is given by:
\begin{equation}
    S = \frac{1}{(1-f)+\frac{f}{s}}
\end{equation}

\section{Properties} 
\label{concurrencyproperties}
We can now examine several useful properties of this speed-up bound $S(N, C)$. These propositions formalize the intuition that removing bottlenecks promotes generalists, never specialists, since only generalists benefit from concurrent access to subtasks. 

\textbf{Proposition 1} (Monotonicity). $S(N,B)$ is non-decreasing in each $C_i$.

\textit{Proof}. $s_i(N,C_i)$ is non‑decreasing in $C_i$ and appears in the denominator of $S$; increasing any $C_i$ weakly decreases the denominator and therefore weakly increases $S$.

\textbf{Proposition 2} (Full concurrency favors generalists). 
Assume \emph{negligible task–switch costs.} If $min_i C_i \ge N$ (every subtask can accommodate all agents), then $S(N,B)=N$ and any throughput‑maximizing policy set satisfies $\mathrm{SI}=0$.

\textit{Proof}. When $C_i\ge N$ for all $i$, we have $s_i(N,C_i)=N$.  Substituting into the definition of $S$ gives us $S(N,B)=1/\!\sum_i f_i/N=N$. 
Because all agents can execute every subtask concurrently, restricting any agent’s action set cannot improve throughput; the optimum is therefore achieved by identical (generalist) policies, for which $\mathrm{SI}=0$.

\textbf{Remark} (Switching costs).
When subtask times vary across workstations or switching between subtasks incurs large delays, the guarantee for generalist policies no longer holds; specialists may achieve higher throughput.

\textbf{Proposition 3} (Bottlenecks induce specialization). 
If $S(N,B)<N$, then no fully generalist team can attain throughput proportional to $N$; a role allocation with $\mathrm{SI}>0$ is required for optimality.

\textit{Proof}. $S<N$ implies \(\exists\,i\) with $C_i<N$.
Under identical generalist policies, the expected number of agents attempting subtask $i$ at any instant is $N$, but at most $C_i$ can make progress; the remainder are idle.  This idle fraction reduces realized speed‑up below $N$.  
A specialist allocation that assigns at most $C_i$ agents to subtask $i$ and redistributes the rest to other subtasks eliminates that idle time.  
Hence any throughput‑optimal policy set must have \(\mathrm{SI}>0\).

\label{DAG_Overcooked}
\begin{figure}[htbp]
    \centering
    \includegraphics[width=0.4\textwidth]{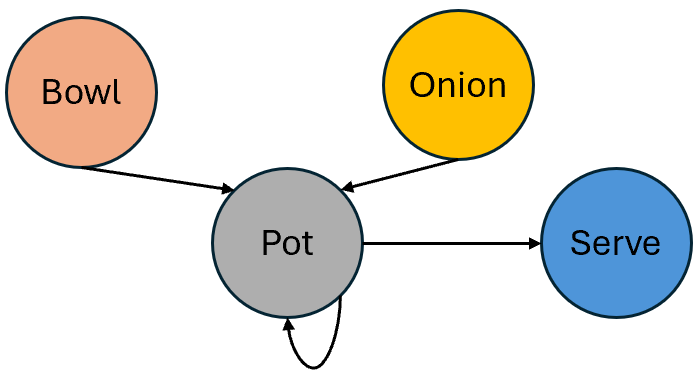}
    \caption{Example of a \textbf{task graph} for making one-onion soup in Overcooked. Task-relevant objects are represented as nodes in a directed acyclic graph (DAG), and edges represent subtasks.}
    \label{fig:DAG}
\end{figure}

\section{Layout Graphs}
\label{layoutgraphs}
A layout graph is an unweighted, directed graph $G=(\mathcal{V},\mathcal{E})$ where:
\[
\mathcal{V} = \{ u \mid u \in \{(i, j) \mid i \in [1, n], j \in [1, m]\} \setminus \text{obstacles} \}
\]
An edge $(u, v) \in \mathcal{E}$ exists if and only if $u$ and $v$ are \textit{adjacent}, i.e., they share a horizontal or vertical edge in the grid. The type of edge between $u$ and $v$ depends on the type of nodes: (i) If $u$ and $v$ are both \textit{walkable nodes}, there is a bidirectional edge such that $(u, v) \in \mathcal{E}$ and $(v, u) \in \mathcal{E}$. (ii) If $v$ is a \textit{task-relevant object} (e.g., workstation) there is a unidirectional edge $(u, v) \in \mathcal{E}$, but $(v, u) \notin \mathcal{E}$, reflecting the fact that agents can interact with but not move through it.

We represent spatial and resource bottlenecks as two global sets over $G=(\mathcal{V}, \mathcal{E})$: $C^s \subseteq \mathcal{E}$, $C^r \subseteq \mathcal{V}$, where $C^s$ consists of edges in $\mathcal{E}$ that constrain movement and $C^r$ consists of nodes in $\mathcal{V}$ with task-relevant objects (e.g., workstations). Each subtask $i$ is constrained by a subset of bottlenecks relevant to the path ($C^s_i$) and resources ($C^r_i$) required to complete it. 

To quantify spatial bottlenecks, we compute \textbf{edge betweenness centrality}, a graph-theoretic measure of congestion where high-centrality edges indicate critical bottlenecks in connectivity \cite{barrat2004architecture}. The betweenness centrality of an edge $e$ in graph $G=(\mathcal{V},\mathcal{E})$ is:
    \begin{equation}
        B(e) = \sum_{s \neq t \in V} \frac{\phi(s,t|e)}{\phi(s,t)},
    \end{equation}
where $\phi(s,t)$ is the number of shortest paths between nodes $s$ and $t$, and $\phi(s,t|e)$ is the number of those paths that pass through edge $e$. Since not all bottlenecks affect every subtask, we compute a per-subtask \textbf{spatial bottleneck score} by summing the centralities of all edges along the shortest path required for that subtask: $C^s_i = \sum_{e \in \text{path}(u,v)} B(e)$,
where path$(u,v)$ is the shortest path between subtask-relevant nodes $u$ and $v$. A higher $C^s_i$ indicates stronger spatial constraints that may inhibit parallel agent movement.

To quantify resource bottlenecks, we simply sum the number of task-relevant nodes that are required for a subtask $i$. For example, in a manufacturing system there may be only two welding stations; thus, only two agents may complete that subtask at the same time. Formally, for each subtask $i$, we define a \textbf{resource bottleneck score} $C^r_i$ as: $C^r_i = \sum_{v \in C^r_i}c(v)$,
where $c(v)$ is the capacity of each task-relevant node (or the maximum agents that can use it at once). 

\section{Training Procedures}
\label{training}
For each trial, we trained two independent agents to play SMAC, MPE, or Overcooked collaboratively through self-play. Our implementation builds upon the PureJaxRL and JaxMARL frameworks for Proximal Policy Optimization \cite{schulman2017proximal, lu2022discovered, flair2023jaxmarl}. Each SMAC and MPE agent was implemented using a recurrent actor-critic network. Observations were embedded using a fully connected ReLU layer, followed by a gated recurrent unit (GRU) to model temporal dependencies. Each Overcooked-AI agent was implemented using an soft actor-critic network \cite{haarnoja2018soft}, in which both the actor and critic networks were composed of two fully-connected layers with 64 hidden units each. We trained each agent using Independent PPO with an Adam optimizer and a learning rate of $2.5 \times 10^{-3}$ for $1 \times 10^7$ timesteps \cite{zhang2018improved}. 

For SMAC, at each timestep, each agent received an observation vector containing local features about itself and other units within its field of view. This included its own health, position, unit type, and weapon cooldown, as well as relative information about nearby allies and enemies (e.g., health, positions, last movement direction, and attack status). The global state included information about all units on the battlefield and was only used during training. The policy produced a categorical distribution over a discrete set of actions, including movement, attack actions, and wait.

For MPE, at each timestep, each agent received a continuous observation vector that encoded its own velocity and position, the relative positions of other agents, and the relative positions of static landmarks. Agents operated in a continuous 2D space. The policy output a categorical distribution over these actions, with masking used when necessary (e.g., to prevent collisions or enforce environment constraints).

For Overcooked-AI, at each timestep, the agents observed the state of the current environment as a $\text{width} \times \text{height} \times 27$-dimensional grid. The first two dimensions corresponded to the spatial coordinates of the grid, while the 27-dimensional vector encoded the state of each grid cell, indicating whether it contained an agent, onion, tomato, bowl, or workstation. The policy produced a categorical distribution over discrete actions.

To accelerate learning in the sparse-reward environment of Overcooked-AI, we applied reward shaping to augment raw rewards with an auxiliary-shaped signal. During reward shaping, each agent received rewards for performing specific actions aligned with the given recipe's requirements: adding onions or tomatoes to the pot when an additional ingredient is needed, cooking the pot once it contains the correct quantity of ingredients, retrieving the soup, and finally serving it. The shaping reward $r_s$ was annealed over a predefined horizon ($2.5 \times 10^{6}$) using a linear schedule. 

\paragraph{Compute}
All experiments were run on A100 GPUs, totaling 1,569 GPU-hours across 17,813 runs. Experiment 1 used 93 GPU-hours (160 runs on MPE and SMAC). Experiment 2 used 120 GPU-hours (1,440 runs), Experiment 3 used 1,333 GPU-hours (16,000 runs), and Experiment 4 used 23 GPU-hours (280 runs).


\section{Experiment 2: Results}
\label{experiment2_app}

\begin{table}[h]
    \centering
    \begin{tabular}{@{}lccc@{}}
        \hline
        \textbf{Number of Pots} & \textbf{Mean SI} & \textbf{Mean Reward} \\ 
        \hline
        1 & 0.676053 & 14.861111\\  
        2 & 0.280481 & 18.867647 \\  
        \hline
    \end{tabular}
    \caption{Average specialization (SI) with varying numbers of pots in Experiment 2.}
    \label{tab:my_label}
\end{table}

\begin{table}[h]
    \centering
    \begin{tabular}{@{}lccc@{}}
        \hline
        \textbf{Spatial Layout} & \textbf{Mean SI} & \textbf{Mean Reward} \\ 
        \hline
        Open & 0.3320525 & 18.7083330 \\  
        Horizontal Divider & 0.332101 & 18.0416668 \\  
        Vertical Divider & 0.318273 & 18.0000000 \\  
        Vertical Space & 0.404964 & 18.0833335 \\
        Long Divider & 0.752011 & 12.2083334 \\
        Two Dividers & 0.2988275 & 17.6666666 \\
        \hline
    \end{tabular}
    \caption{Average specialization (SI) in various spatial layouts in Experiment 2.}
    \label{tab:my_label}
\end{table}

\section{Experiment 3: Significance Tests}
\label{experiment1sigtests}
\textbf{Number of Pots and Workstations.} 
First, we found a significant effect of the number of pots, but not the total number of workstations, on specialization. Agents tended to specialize more with one pot ($M=0.77, SD=0.30$) than two ($M=0.57, SD=0.41$) but not when presented with more of other workstations (e.g., bowls, serving stations). A Welch's t-test between layouts with one pot versus two revealed a significant difference in SI, $t(2842.72)=14.98, p=7.23 \times 10^{-49}$. Conversely, a one-way ANOVA showed a non-significant difference in SI across the full range of possible workstation combinations ($F(5, 3094) = 0.82, p = 0.54$). 

\textbf{Spatial Bottlenecks.} Next, we found a significant effect of spatial bottlenecks on specialization. Agents specialized less in layouts with fewer bottlenecks ($M=0.61, SD=0.38$) than in those with medium ($M=0.69, SD=0.37$) or high bottlenecks ($M=0.86, SD=0.27$). Using edge betweenness centrality, we quantified the average extent of spatial bottlenecks across the entire layout. Betweenness values were divided into three equal-width bins: Low $[2.110, 3.194)$, Medium $(3.195, 4.277]$, and High $(4.277, 5.360]$. Smaller centrality values represented a fewer number of nodes that were critical to many paths in the layout graph, corresponding to a smaller number of spatial bottlenecks. Larger centrality values represented more critical nodes, corresponding to greater spatial bottlenecks. A one-way ANOVA showed a significant difference in SI across bins ($F(2, 2097) = 34.10, p = 2.26 \times 10^{-15}$). Pairwise comparisons using Tukey's HSD showed that SI was significantly higher with medium centrality than low ($p = 3.18 \times 10^{-7}$), high centrality than medium ($p = 1.07 \times 10^{-12}$), and high centrality than low ($p = 3.82 \times 10^{-20}$). 

\textbf{Recipe Complexity.} Finally, we found a significant effect of recipe complexity (the number of ingredients required to make a soup) on SI. Agents tended to specialize less when recipes had one ingredient ($M=0.54, SD=0.47$) than with two ($M=0.73, SD=0.35$) or three ($M=0.67, SD=0.32$). A one-way ANOVA showed a significant difference in SI across recipes ($F(2,3097) = 57.22, p = 3.95 \times 10^{-25}$). Pairwise comparisons using Tukey's HSD revealed that SI was significantly higher for recipes with two ingredients compared to one ($p=4.46 \times 10^{-19}$) and recipes with three ingredients compared to one ($p=4.37 \times 10^{-10}$), but significantly lower for recipes with three ingredients compared to two ($p=7.77 \times 10^{-6}$). 

\section{Experiment 3: Evaluation}
\label{expt1_regression}
For each of the 14 layouts, there were a total of 16 workstation combinations with three randomized positions each. We computed the predicted parallelizability, $S$, for each of these combinations and averaged over the three randomized locations to compute a measure of task parallelizability for each combination of layout, workstations, and recipe.

We also fit a logistic regression model where the binary outcome variable was low versus high specialization (SI $\geq 0.5$) and the predictor variable was task parallelizability. To address the class imbalance in the SI distribution, we used class-weight balancing to automatically adjust class weights inversely proportional to their frequencies in the training data. 

\begin{figure*}[h]
     \centering
     \includegraphics[width = 0.9\linewidth]{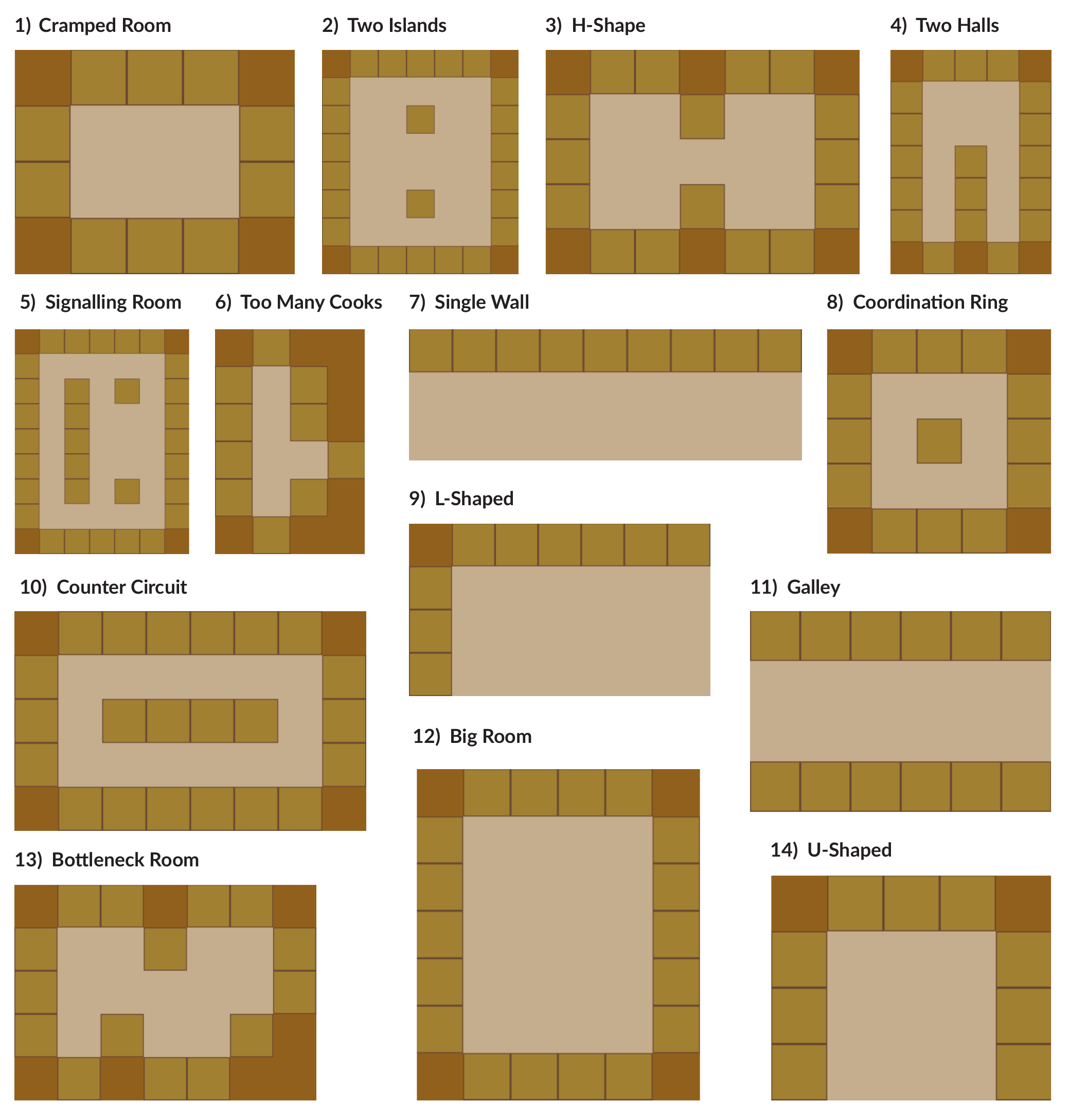}
    \caption{Overcooked Spatial Layouts in Experiment 3.}
    \label{fig:layouts}
\end{figure*}

\begin{figure*}[h]
     \centering
     \includegraphics[width = 0.9\linewidth]{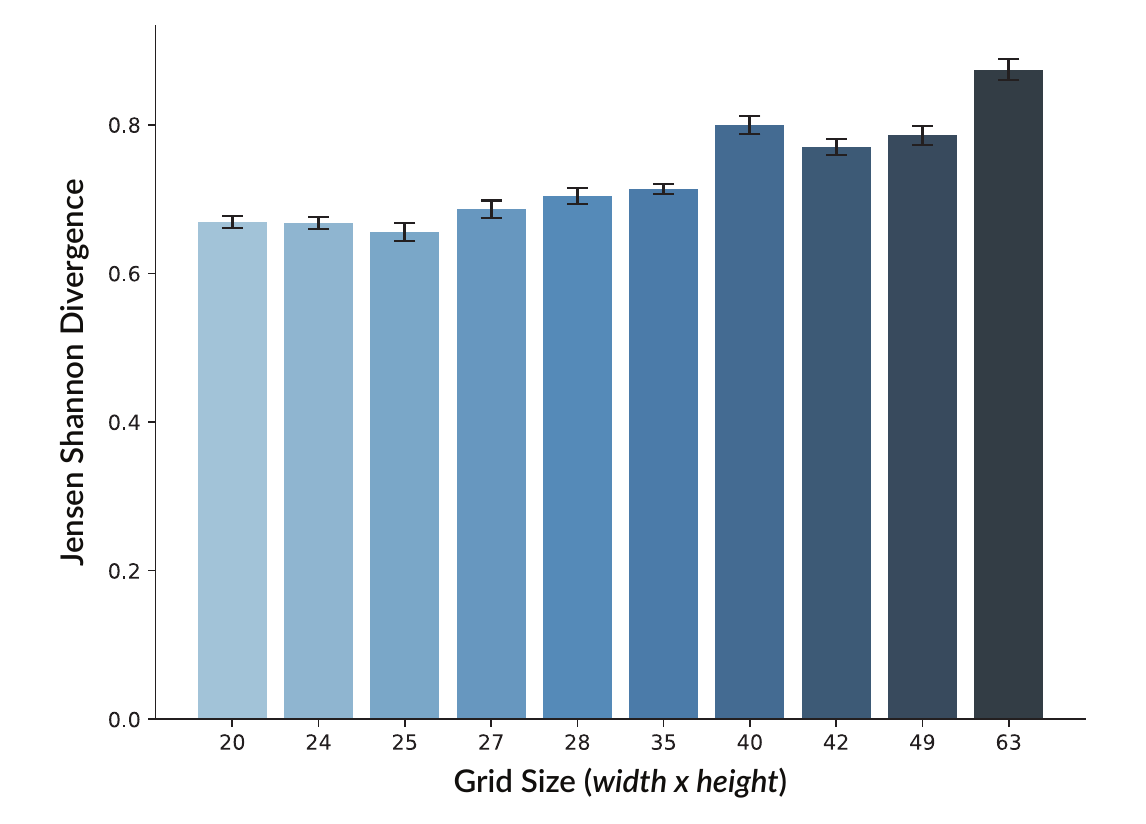}
    \caption{Effect of grid size on observed specialization (SI) in Experiment 3. As the size of the grid increased, agends tended to specialize more (even when presented with the same number of workstations and bottlenecks). }
    \label{fig:gridsize}
\end{figure*}

\begin{table}[h]
    \centering
    \begin{tabular}{@{}lccc@{}}
        \hline
        \textbf{Spatial Layout} & \textbf{Mean SI} & \textbf{Mean Reward} \\ 
        \hline
        H\_shape & 0.666725 & 14.165217 \\  
        L\_shaped & 0.599760 & 14.050847 \\ 
        U\_shaped & 0.588023 & 15.222689 \\ 
        big\_room & 0.732749 & 13.207627 \\
        bottleneck\_room & 0.686606 & 14.212121 \\
        coord\_ring & 0.587970 & 14.845188 \\
        counter\_circuit & 0.829034 & 12.990385 \\
        cramped\_room & 0.597489 & 15.879167 \\
        gallery & 0.603094 & 14.564854 \\
        semi\_forced\_coord & 0.744749 & 13.440191 \\
        signaling\_room & 0.860218 & 12.813953 \\
        single\_wall & 0.628655 & 12.808696 \\
        too\_many\_cooks & 0.631687 & 14.099567 \\
        two\_islands & 0.758321 & 13.133005 \\
        \hline
    \end{tabular}
    \caption{Average specialization (SI) observed in different spatial layouts in Experiment 3.}
    \label{tab:experiment1_layouts}
\end{table}

\begin{table}[h]
    \centering
    \begin{tabular}{@{}lcc@{}}
        \hline
        \textbf{Workstations} & \textbf{Mean SI} & \textbf{Std SI} \\
        \hline
        1 onions & 0.69 & 0.38 \\
        2 onions & 0.66 & 0.40 \\
        1 pots & 0.76 & 0.33 \\
        2 pots & 0.58 & 0.42 \\
        1 serving & 0.66 & 0.39 \\
        2 serving & 0.69 & 0.38 \\
        0 tomatoes & 0.60 & 0.41 \\
        1 tomatoes & 0.79 & 0.31 \\
        1 bowls & 0.68 & 0.38 \\
        2 bowls & 0.67 & 0.39 \\
        \hline
    \end{tabular}
    \caption{Mean and standard deviation of observed specialization (SI) for varying numbers of workstations in Experiment 3.}
    \label{tab:quantities_SI}
\end{table}

\begin{table}[h]
    \centering
    \begin{tabular}{@{}lcc@{}}
        \hline
        \textbf{Recipes} & \textbf{Mean SI} & \textbf{Std SI} \\
        \hline
        1 onion & 0.56 & 0.48 \\
        2 onions & 0.65 & 0.39 \\
        3 onions & 0.59 & 0.36 \\
        1 onion 1 tomato & 0.82 & 0.33 \\
        2 onions 1 tomato & 0.77 & 0.29 \\
        \hline
    \end{tabular}
    \caption{Mean and standard deviation observed specialization (SI) for different recipes in Experiment 3.}
    \label{tab:recipes_SI}
\end{table}

\begin{table}[h]
\label{table1_correlations}
\centering
\caption{Correlations Across Spatial Layouts}
\begin{tabular}{@{}lccc@{}}
\toprule
\textbf{Base Layout}        & \textbf{Correlation} & \textbf{p-value} \\ \midrule
H\_shape                    & -0.493              & 0.004                          \\
L\_shaped                   & -0.699              & $< 0.001$                    \\
U\_shaped                   & -0.679              & $< 0.002$                    \\
big\_room                   & -0.568              & 0.001                          \\
bottleneck\_room            & -0.544              & 0.001                          \\
coord\_ring                 & -0.307              & 0.087                          \\
counter\_circuit            & -0.361              & 0.042                          \\
cramped\_room               & -0.619              & $< 0.001$                       \\
gallery                     & -0.524              & 0.002                          \\
semi\_forced\_coord         & -0.555              & 0.001                          \\
signaling\_room             & -0.479              & 0.009                           \\
single\_wall                & -0.553              & 0.001                           \\
too\_many\_cooks            & -0.272              & 0.132                           \\
two\_islands                & -0.244              & 0.178                           \\ \midrule
\textbf{Average Correlation} & \textbf{-0.494}     &                  &                      \\ \bottomrule
\end{tabular}
\caption{Correlations between predicted task parallelizability ($S$) and observed specialization (SI) across varying spatial layouts in Experiment 3. The significance of this relationship varied across layouts, especially those with different grid sizes, prompting the controlled design of Experiments 2 and 3.}
\end{table}

\begin{figure}[htp]
    \centering
    \includegraphics[width = \linewidth]{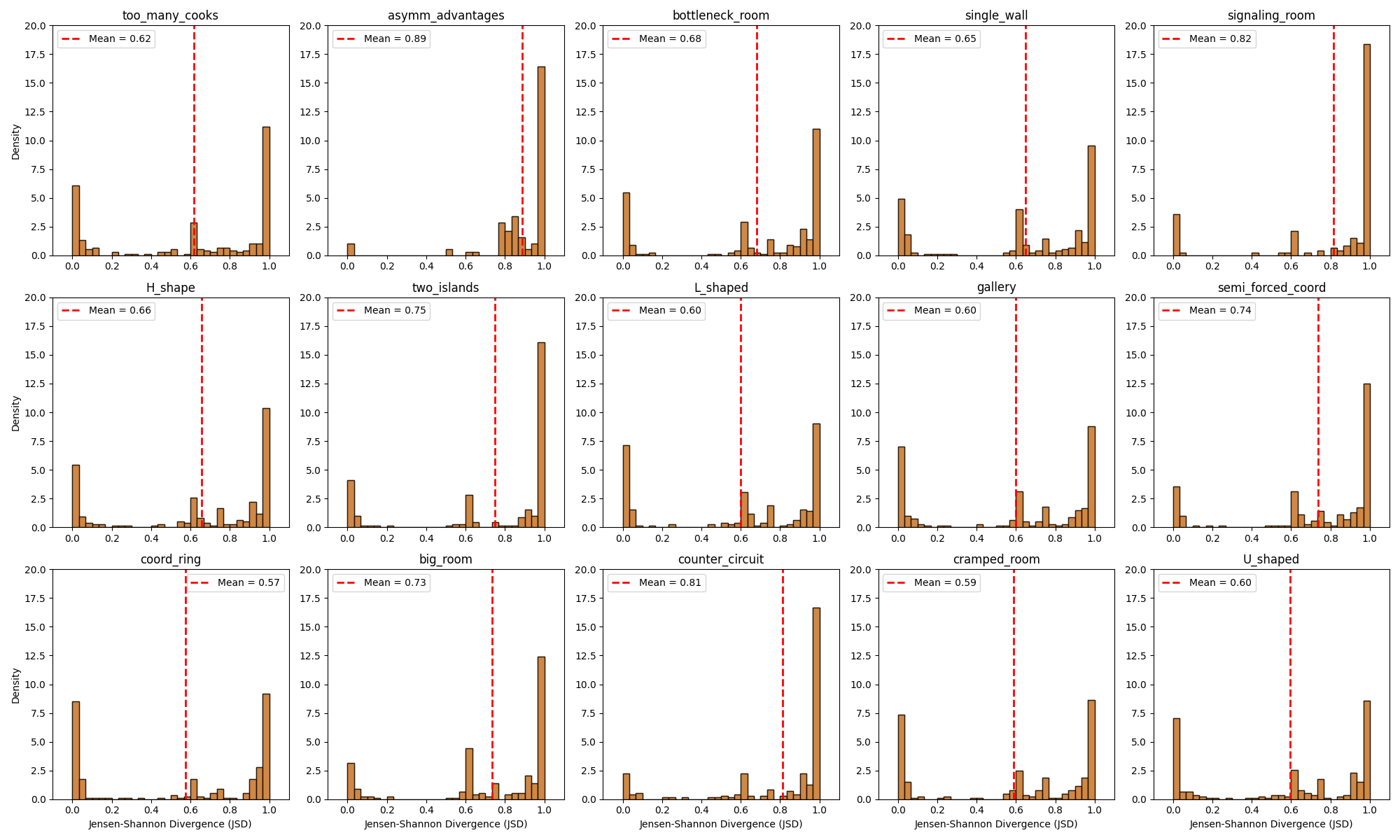}
    \caption{Specialization (SI) distributions across all permutations in Experiment 3. Each subfigure represents the SI distribution grouped by a specific spatial layout.}
    \label{fig:SI_layout}
\end{figure}

\begin{figure}[htp]
    \centering
    \includegraphics[width = \linewidth]{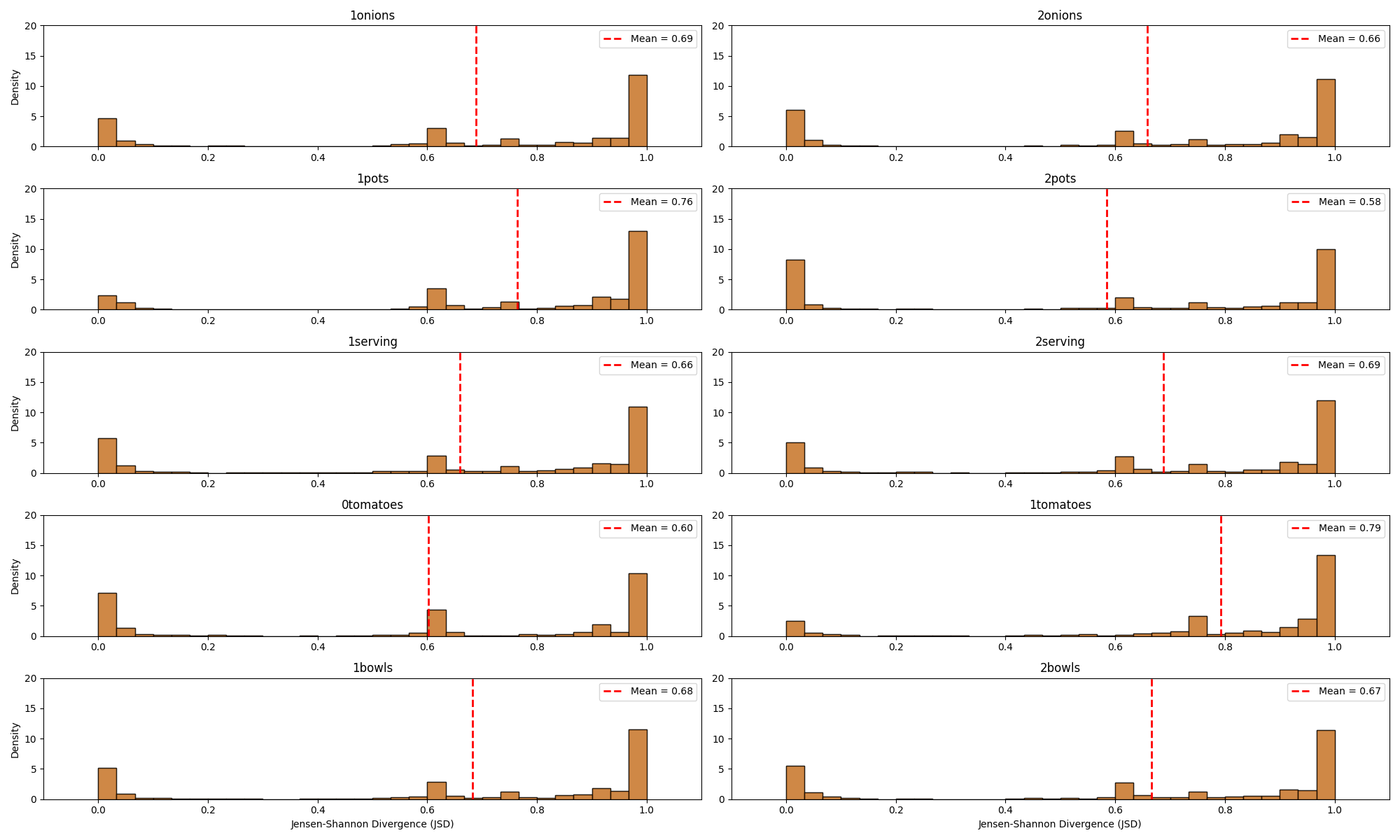}
    \caption{Specialization (SI) distributions across all permutations in Experiment 3. Each subfigure represents the SI distribution for a specific number of workstations.}
    \label{fig:SI_layout}
\end{figure}

\begin{figure}[htp]
    \centering
    \includegraphics[width = \linewidth]{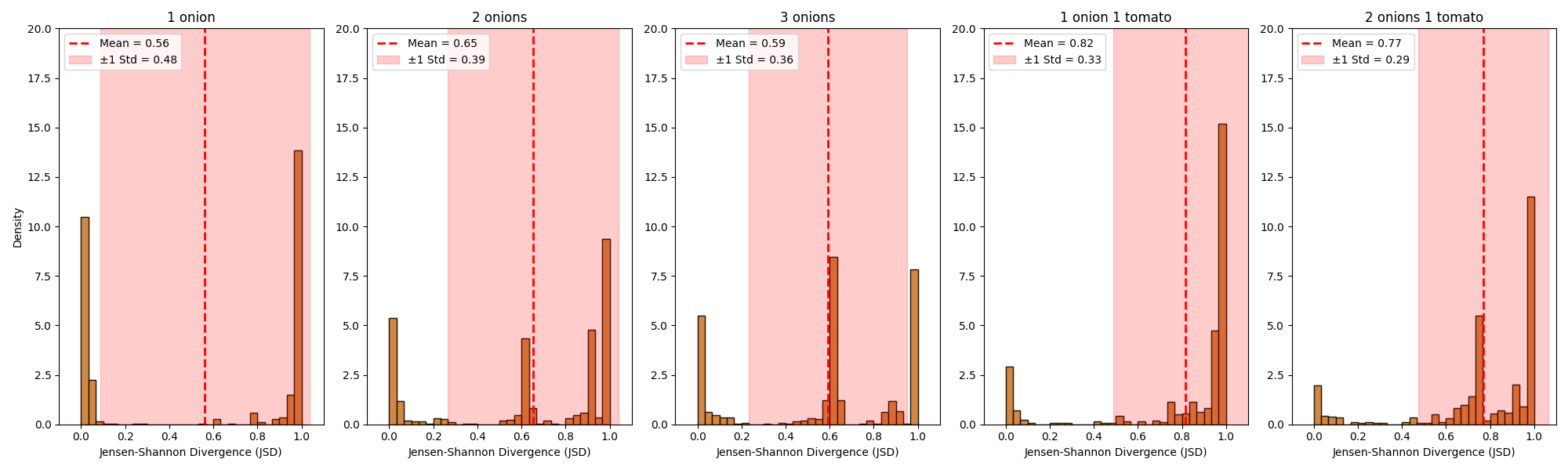}
    \caption{Specialization (SI) distributions across all permutations in Experiment 3. Each subfigure represents the SI distribution grouped by a specific recipe.}
    \label{fig:SI_layout}
\end{figure}

\begin{figure}[htp]
    \centering
    \includegraphics[width = \linewidth]{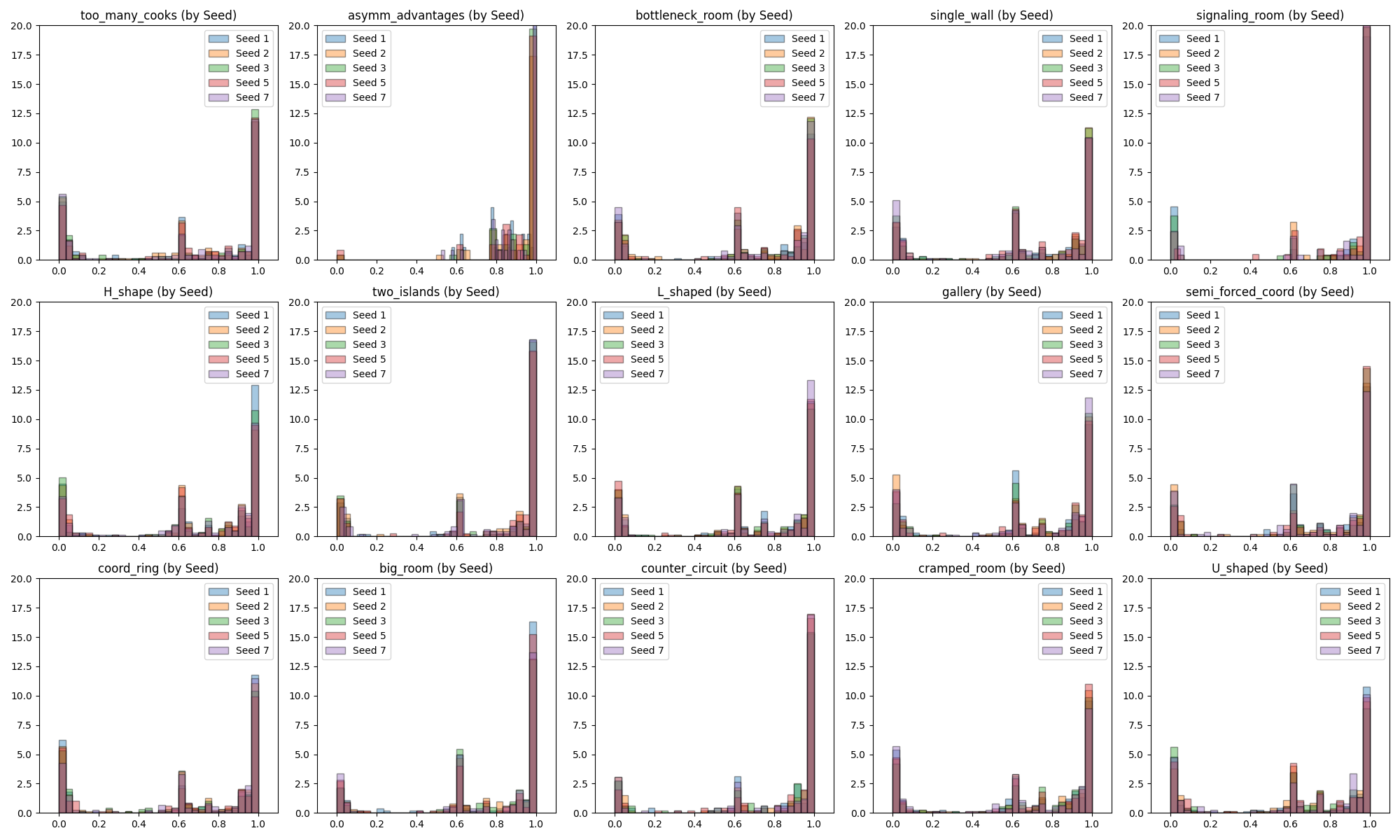}
    \caption{Specialization (SI) distributions across all random seeds in Experiment 3. Each subfigure represents the SI distribution grouped by a specific layout. The distributions of strategies appear mostly consistent across different random seeds. Five random seeds were used: 1, 2, 3, 5, and 7.} 
    \label{fig:SI_layout}
\end{figure}

\begin{figure}[htp]
    \centering
    \includegraphics[width = \linewidth]{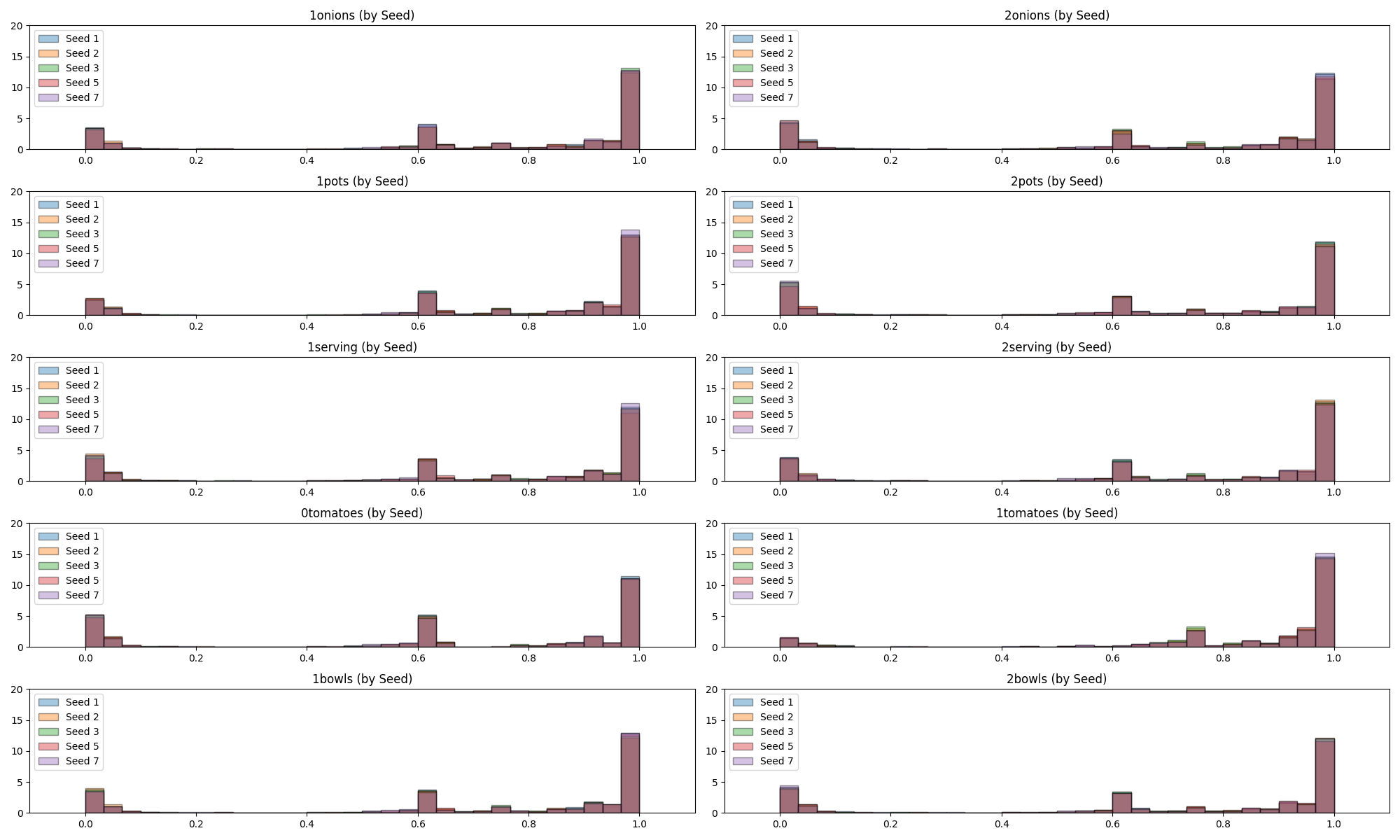}
    \caption{Specialization (SI) distributions across all random seeds in Experiment 3. Each subfigure represents the SI distribution grouped by the number of workstations. The distributions appear consistent across the different random seeds. Five random seeds were used: 1, 2, 3, 5, and 7.} 
    \label{fig:SI_layout}
\end{figure}

\begin{figure}[htp]
    \centering
    \includegraphics[width = \linewidth]{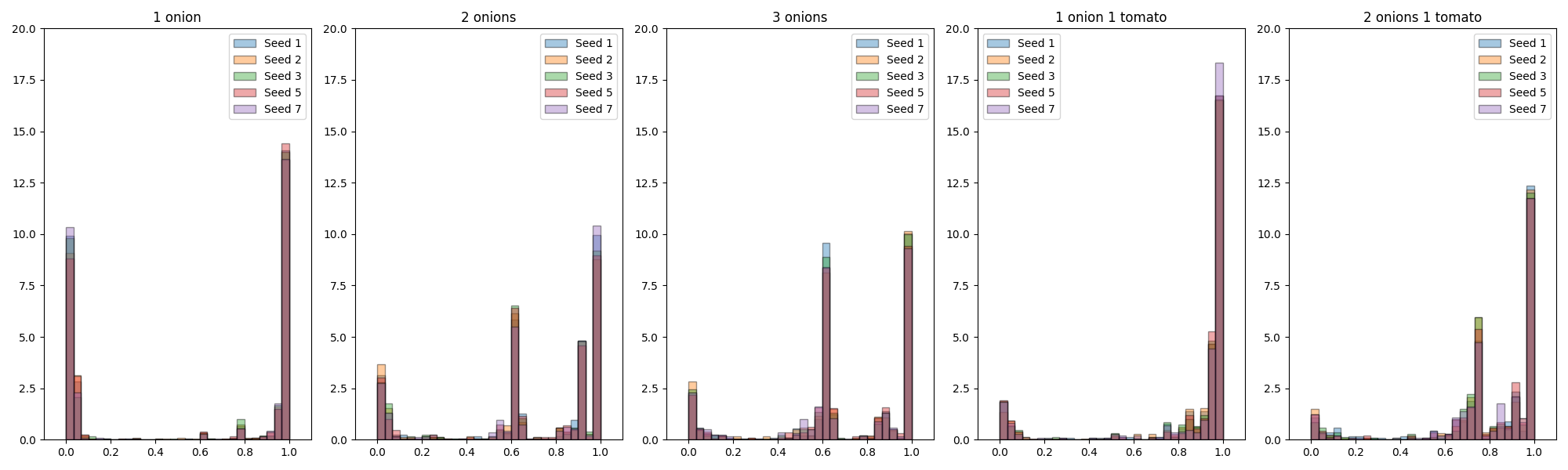}
    \caption{Specialization (SI) distributions across all random seeds in Experiment 3. Each subfigure represents the SI distribution grouped by a specific recipe. The distributions appear consistent across the different random seeds. Five random seeds were used: 1, 2, 3, 5, and 7.} 
    \label{fig:SI_layout}
\end{figure}



\section{Experiment 4: Results}


\setlength{\tabcolsep}{2pt}

\begin{table}[ht]
\centering
\label{tab:summary_stats}
\begin{small}
\begin{tabular}{ccccccccc}
\toprule
\textbf{Layout} & \textbf{Experiment} & \textbf{Recipe} & \textbf{Mean SI} & \textbf{Std SI} & \textbf{Count} & \textbf{Mean Reward} & \textbf{Std Reward} & \textbf{Reward Count} \\
\midrule
1 & A & 1-Onion & 0.750219 & 0.397342 & 10 & 24.8 & 6.014797 & 10 \\
1 & A & 3-Onion & 0.307466 & 0.477938 & 10 & 13.7 & 3.301515 & 10 \\
1 & B & 1-Onion  & 0.750219 & 0.397342 & 10 & 24.8 & 6.014797 & 10 \\
1 & B & 3-Onion & 0.307466 & 0.477938 & 10 & 13.7 & 3.301515 & 10 \\
2 & A & 1-Onion  & 0.570917 & 0.481485 & 10 & 24.0 & 4.136558 & 10 \\
2 & A & 3-Onion & 0.712140 & 0.355738 & 10 & 9.8  & 3.047768 & 10 \\
2 & B & 1-Onion  & 0.751195 & 0.431745 &  9 & 27.8 & 7.052186 & 10 \\
2 & B & 3-Onion & 0.991283 & 0.013504 &  6 & 8.4  & 0.516398 & 10 \\
3 & A & 1-Onion & 0.683808 & 0.453647 & 10 & 27.2 & 5.329165 & 10 \\
3 & A & 3-Onion & 0.428280 & 0.458991 &  9 & 9.5  & 4.478343 & 10 \\
3 & B & 1-Onion & 0.579108 & 0.497131 & 10 & 28.9 & 6.136412 & 10 \\
3 & B & 3-Onion & 0.887113 & 0.197573 & 10 & 8.7  & 0.483046 & 10 \\
4 & A & 1-Onion & 0.956536 & 0.130393 &  9 & 23.2 & 8.202980 & 10 \\
4 & A & 3-Onion & 0.686612 & 0.392708 &  6 & 8.0  & 3.829708 & 10 \\
4 & B & 1-Onion & 0.462766 & 0.497691 & 10 & 31.3 & 4.922736 & 10 \\
4 & B & 3-Onion & 0.657041 & 0.433589 &  5 & 8.2  & 3.794733 & 10 \\
5 & A & 1-Onion & 0.806959 & 0.310521 & 10 & 24.2 & 3.119829 & 10 \\
5 & A & 3-Onion & 0.851840 & 0.235864 &  6 & 6.7  & 5.355164 & 10 \\
5 & B & 1-Onion & 0.694087 & 0.476603 & 10 & 28.9 & 4.931757 & 10 \\
5 & B & 3-Onion & 0.478804 & 0.391161 &  6 & 7.9  & 3.348300 & 10 \\
6 & A & 1-Onion & 0.975668 & 0.072995 &  9 & 24.2 & 3.293090 & 10 \\
6 & A & 3-Onion & 0.899543 & 0.210584 &  7 & 8.3  & 0.483046 & 10 \\
6 & B & 1-Onion & 0.504944 & 0.451051 & 10 & 26.4 & 11.393956& 10 \\
6 & B & 3-Onion & 0.520999 & 0.319708 &  7 & 9.4  & 2.988868 & 10 \\
7 & A & 1-Onion & 1.000000 & 0.000000 &  8 & 24.3 & 3.335000 & 10 \\
7 & A & 3-Onion & 0.750629 & 0.354937 &  8 & 6.7  & 3.560587 & 10 \\
7 & B & 1-Onion & 0.583097 & 0.488816 &  8 & 26.3 & 7.424434 & 10 \\
7 & B & 3-Onion & 0.713627 & 0.403798 &  5 & 6.5  & 3.439961 & 10 \\
\bottomrule
\end{tabular}
\end{small}
\caption{Detailed results from Experiment 4. Each row represents a unique combination of spatial layout, experiment, and recipe, showing the mean, standard deviation (std), and count (number of runs) for both specialization (SI) and reward metrics. Experiment A corresponds to expanding exploration, while Experiment B denotes constant exploration. The SI count is less than 10 for certain permutations because the policy failed to converge, resulting in the agents taking no meaningful actions.}
\end{table}


\setlength{\tabcolsep}{10pt}

\begin{table}[t]
\centering
\label{tab:corr_aggregated}
\begin{tabular}{lcccc}
\toprule
\textbf{Recipe}     & \textbf{Exploration} & \textbf{Metric}  & \textbf{Corr.}      & \textbf{p-value} \\
\midrule
1-Onion             & Expanding            & SI              & 0.7924          & 0.0336 \\
                    &                     & Reward           & -0.2512         & 0.5869 \\
                    & Constant             & SI              & -0.5850         & 0.1677 \\
                    &                     & Reward           & 0.0609          & 0.8969 \\
\midrule
3-Onion             & Expanding            & SI              & 0.7549          & 0.0498 \\
                    &                     & Reward           & -0.8556         & 0.0140 \\
                    & Constant             & SI              & -0.0423         & 0.9283 \\
                    &                     & Reward           & -0.6947         & 0.0832 \\
\bottomrule

\end{tabular}
\caption{Correlations from Experiment 4. Correlations between specialization (SI) and reward for one-onion and three-onion Soup recipes under expanding and constant exploration conditions. The table reports correlations and p-values ($p$) for SI and reward in relation to layout size.}
\end{table}

\begin{table}[h]
\centering
\label{tab:ttests}
\begin{tabular}{llll}
\toprule
\textbf{Condition} & \textbf{Metric} & \textbf{t-value} & \textbf{p-value} \\
\midrule
\multicolumn{4}{c}{\textbf{Combined Recipes}} \\
layout\_1 vs layout\_7 & SI & -2.229 & 0.029 \\
layout\_0 vs layout\_7 & Reward & 0.506 & 0.614  \\
\midrule
\multicolumn{4}{c}{\textbf{1-Onion Soup Recipe}} \\
layout\_1 vs layout\_7 & SI & -0.315 & 0.755  \\
layout\_1 vs layout\_7 & Reward & -1.278 & 0.210  \\
\midrule
\multicolumn{4}{c}{\textbf{3-Onion Soup Recipe}} \\
layout\_1 vs layout\_7 & SI & -2.820 & 0.008 \\
layout\_1 vs layout\_7 & Reward & 5.812 & 0.000 \\
\bottomrule
\end{tabular}
\caption{Results from Experiment 4. Pairwise T-tests comparing the smallest (\textit{Layout 1}) layout with the largest layout (\textit{Layout 67}).  }
\end{table}

\end{document}